\documentclass[preprint]{elsarticle}

\usepackage{lineno,hyperref}
\usepackage{amsmath,amsfonts,amssymb} 
\usepackage{epsfig,color,graphicx}
\usepackage{xcolor}
\newcommand{\e}{\varepsilon}
\newcommand{\g}{\gamma}

\newcommand{\w}{\omega}
\modulolinenumbers[5]

\journal{Physica D}

\begin{document}

\begin{frontmatter}

\title{The Kuramoto model of coupled oscillators with a bi-harmonic coupling
function}

\author[Potsdam,Nizhny]{M. Komarov\corref{mycorrespondingauthor}}
\cortext[mycorrespondingauthor]{Corresponding author}
\ead{maxim.a.komarov@gmail.com}

\author[Potsdam,Nizhny]{A. Pikovsky}

\address[Potsdam]{Department of Physics and Astronomy, Potsdam University,
Karl-Liebknecht-Str 24/25, 14476 Potsdam, Germany, 
}
\address[Nizhny]{Department of Control Theory, Nizhny Novgorod State University,
Gagarin Avenue 23, 603950 Nizhny Novgorod, Russia}

\begin{abstract}
We study synchronization in a Kuramoto model of 
globally coupled phase oscillators 
with a bi-harmonic
coupling function, in the thermodynamic limit of large populations.
We develop a method for an analytic
solution of self-consistent equations describing uniformly rotating 
complex order 
parameters, both for single-branch (one possible
state of locked oscillators) and multi-branch (two possible values of locked phases)
entrainment.  We show that synchronous states coexist with the neutrally 
linearly stable asynchronous regime. The latter has a finite life time for 
finite ensembles, this time grows with the ensemble size as a power law. 
\end{abstract}

\begin{keyword}
Kuramoto model\sep Bi-harmonic coupling function\sep Multi-branch entrainment \sep
Synchronization
\end{keyword}

\end{frontmatter}


\section{Introduction}

Large systems of coupled nonidentical oscillators are of general interest in
various
branches of science. They describe Josephson 
junction circuits~\citep{Wiesenfeld-Swift-95,Wiesenfeld-Colet-Strogatz-96,
Wiesenfeld-Colet-Strogatz-98}, 
electrochemical~\citep{Kiss-Zhai-Hudson-02a} and
spin-torque~\citep{Grollier-Cros-Fert-06,georges:232504} oscillators, 
as well as variety of interdisciplinary applications including pedestrian
induced oscillations of footbridges~\citep{Eckhardt_et_al-07}, applauding
persons~\citep{Neda_etal-00}, and others. 
Similar models are also used in biology, for example in studying of neural
ensembles dynamics~\citep{Golomb-Hansel-Mato-01,%
Breakspear-Heitmann-Daffertshofer-10} and
systems describing circadian clocks in mammals~\citep{Gonze-05, Bordyugov-13}.
In many cases the analysis of large ensembles consisting of heterogeneous
oscillators can be
successfully performed in the phase approximation~\citep{Kuramoto-84,
Pikovsky-Rosenblum-Kurths-01}. Indeed, if the interaction between the 
elements is weak, the amplitudes are enslaved, and
the dynamics of self-sustained oscillators can be effectively described by
a relatively simple system of coupled phase equations. 
The special case of a globally coupled network of phase oscillators (so-called
Kuramoto model~\citep{Kuramoto-84,Kuramoto-75}) attracted a lot of
attention~\citep{Acebron-etal-05} and has been established as a paradigmatic
model describing transitions from incoherent to synchronous states in the
ensembles of coupled oscillators.

Quite a complete analysis of the Kuramoto model can be performed 
in the case of a
harmonic
sin-coupling 
function~\cite{Kuramoto-84,Ott-Antonsen-08,Ott-Antonsen-09}, although even here
non-trivial scenaria of transition to synchrony have been 
reported~\cite{Omelchenko-Wolfrum-12}.
Less studied is the case of more general coupling functions, 
containing many harmonics.
Here we perform a systematic study of the synchronous regimes 
for a bi-harmonic coupling function
(see~\cite{Komarov-Pikovsky-13a} for a short presentation of these results which have been later confirmed in \citep{Li-Ma-Li-Yang-14}). 
We introduce the model
and discuss previous findings in Section~\ref{sec:KMBHC}. Then in 
Section~\ref{sec:sca} we give a general solution of the self-consistent 
equations describing rotating-wave synchronous solutions.
In Section~\ref{sec:sym} we give a detailed analysis of the simplest 
symmetric case (no phase shifts in the coupling), while a general situation is 
illustrated
in Section~\ref{sec:asym}. In Conclusion we summarize the results and outline 
open questions.
In this paper we focus on the deterministic oscillator dynamics, the case of 
noisy oscillators will be considered 
elsewhere~\cite{Komarov-Pikovsky-unpublished}.

\section{Kuramoto Model and Bi-Harmonic Coupling}
\label{sec:KMBHC}

The general Kuramoto model is formulated as a system of differential equations for
the phases $\phi_k$ of $N$ oscillators:
\begin{equation}
\dot{\phi_k} =
\omega_k+\frac{1}{N}\sum_{n=1}^{N}\Gamma(\phi_n-\phi_k),\qquad k=1,..,N.
\label{eq:km}
\end{equation}
All the oscillators are identical, except for diversity of
the natural frequencies $\omega_k$, distributed according to a certain 
distribution function $g(\omega)$.
The level of coherence in the network of phase oscillators can be essentially
described by order parameters $R_n$ defined by:
$$
R_n e^{i\Theta_n} = \frac{1}{N}\sum_{k=1}^{N} e^{in\psi_k},\ n\in\mathbb{N}.
$$
The state with $R_n=0$ for all $n$ 
corresponds to a 
purely incoherent dynamics (uniform distribution of the phases), 
while non-zero
values of at least some order parameters indicate for certain synchrony in the 
ensemble.
In the case of pure sinusoidal coupling, 
$\Gamma (x)= \varepsilon\sin(x+\alpha)$, the
original analysis by Kuramoto~\citep{Kuramoto-75,Kuramoto-84} and its 
subsequent
extensions~\citep{Sakaguchi-Kuramoto-86,Crawford-95,%
Crawford-Davies-99,Ott-Antonsen-08,
Ott-Antonsen-09} revealed a clear picture of a transition from asynchronous 
state
to coherence in the thermodynamical limit $N\to \infty$. 
It was shown that above certain critical value of the 
coupling ($\e > \e_c$),
the system undergoes a 
transition from disordered behavior to synchronous
collective motion via a supercritical bifurcation with the main order parameter 
obeying $R_1 \sim (\e-\e_c)^{\frac{1}{2}}$. 

The situation is much less trivial for more general coupling functions $\Gamma$.
The presence of higher harmonics in coupling
function~\cite{Daido-96,Crawford-95,Crawford-Davies-99,Chiba-Nishikawa-11} may
change scaling of the order parameter to linear law $R_1 \sim \e-\e_c$. 
Moreover, as has been already mentioned in an 
early paper by Winfree~\citep{Winfree-80} and 
in subsequent numerical
studies by Daido in~\citep{Daido-95,Daido-96a},  sufficiently strong
higher modes in the coupling function $\Gamma$ 
may cause a so-called multibranch entrainment, in which a huge number
of stable or multistable phase-locked states exists. 
In certain cases the interplay between synchronizing action of one coupling
mode and
repelling force from another one can be a reason for an oscillatory behavior of
macroscopic order parameters~\citep{Hansel-93}.

This paper is devoted to a 
systematic study of the Kuramoto model in the case of
a general bi-harmonic coupling function
\begin{equation}
\Gamma(x) = \e\sin(x-\beta_1)+\gamma\sin(2x-\beta_2)
\label{eq:Gamma}
\end{equation}
in the thermodynamic limit $N\to\infty$.
In Section~\ref{sec:sca} we formulate an analytic self-consistent
approach~\citep{Kuramoto-75,Kuramoto-84,Skardal-Ott-Restrepo-11} which allows us
to calculate stationary or uniformly rotating 
order parameters $R_{1,2}$ (including all possible multi-branch
entrainment states) depending on the parameters of the bi-harmonic 
coupling function
$\Gamma$. 
Based on the self-consistent method, we present in 
Section~\ref{sec:sym} a complete diagram 
of uniformly rotating states with constant order parameters, for a
special case of symmetric coupling function $\Gamma$ ($\beta_{1,2}=0$). 
Surprisingly, (i) synchronous solutions appear \textit{prior} to the stability
threshold of incoherent state; (ii)  these regimes 
have order parameters that can take values anywhere in the range $(0,R_{max}]$
for some $R_{max}< 1$; 
(iii) there is a huge multiplicity of these states for fixed coupling parameters
(multi-branch entrainment) which can also appear for \textit{relatively weak}
second mode (when parameter $\gamma$ is small compare to absolute value of $\varepsilon$) in the coupling.
Here we also illustrate the multiplicity of solutions,
and, combining the self-consistent approach and a perturbative
analysis, we derive the scaling laws of $R_{1,2}(\e,\gamma)$ 
near the transition
points where coherent state appears.

For a general case of non-zero $\beta_{1,2}$, 
consideration of the self-consistent 
equations becomes rather tedious due to a large
number of parameters involved.
We restrict our attention in Section~\ref{sec:asym} to several examples with
multibranch entrainment and to already mentioned oscillatory
states~\citep{Hansel-93}.

Before proceeding with the analysis, we mention 
three examples of
realistic
physical systems where the second harmonics term in the coupling function
is strong or even dominating.
The first example is the classical Hyugens' setup with pendulum
clocks suspended on a
common beam (common platform).  
The horizontal 
displacement of the beam leads to the first harmonics coupling
$\sim\varepsilon$, while the vertical
mode produces the second harmonics term
$\sim\gamma$~\citep{Czolczynski_etal-13}. 
We give a derivation of the 
phase equations for the case where both 
horizontal and vertical displacements of the platform are 
present, in Appendix~\ref{sec:appA}, where Eq.~(\ref{eq:app-fin}) is in fact
the Kuramoto model  with bi-harmonic coupling.
Another example are recently experimentally 
realized $\varphi-$Josephson junctions~\citep{Goldobin_etal-11}, 
where the
dynamics of a single junction in the array is governed by a double-well energy
potential.
Therefore one can expect strong effects caused by the second harmonics in 
the interaction. The third example are experiments with globally 
coupled electrochemical
oscillators~\citep{Kiss-Zhai-Hudson-05,Kiss-Zhai-Hudson-06}, where 
a pronounced 
second harmonics has been observed in the coupling function inferred from
the experimental data.

\section{Self-consistent equations and their solution}
\label{sec:sca}

We start our analysis with reformulation of equation (\ref{eq:km})
for the bi-harmonic coupling as
\[
\dot\varphi_k=\omega_k+\varepsilon \text{Im}\left[ e^{-i\beta_1-i\varphi_k}\frac{1}{N}\sum_n e^{i\varphi_n}\right]
+\gamma \text{Im}\left[ e^{-i\beta_2-i2\varphi_k}\frac{1}{N}\sum_n e^{i2\varphi_n}\right]   \;.
\]
 In the
thermodynamical limit, using the two relevant order parameters
$R_{1,2}e^{i\Theta_{1,2}}$, we obtain:
\begin{equation}
\dot{\varphi} = \omega+\varepsilon R_1 \sin(\Theta_1-\varphi-\beta_1)+\gamma R_2
\sin(\Theta_2-2\varphi-\beta_2)      \;.
\label{eq:bkm_th}
\end{equation}
We assume the natural frequencies $\omega$ to be distributed 
according to a symmetric, single-maximum function $g(\omega)$.
In the thermodynamical limit the complex order parameters $R_me^{i\Theta_m}$ can
be represented using the conditional distribution 
function $\rho(\varphi|\omega)$:
\begin{equation}
R_me^{i\Theta_m}= \iint d\varphi d\omega\;g(\w)\rho(\varphi|\omega)
e^{im\varphi},\qquad
m=1,2\;.
\label{eq:opint}
\end{equation}

Let us perform a following transformation of variables to the rotating 
(with some frequency $\Omega$) reference frame:
\begin{equation}
\Theta_1 = \Omega t+\theta_1;\ \Theta_2=\Omega t+\theta_2;\ \varphi = \Omega
t+\theta_1-\beta_1+\psi\;.
\label{eq:var_tr}
\end{equation}
Then equation (\ref{eq:bkm_th}) changes as follows:
\begin{equation}
\dot{\psi} = \omega-\Omega+\e R_1\sin(-\psi)+\gamma R_2 \sin(\theta_2 -
2\theta_1+2\beta_1-\beta_2-2\psi)\;.
\label{eq:bkm_th1}
\end{equation}
It is convenient to introduce a set of parameters 
$\{R,\ u,\ v,\ z\}=\mathbf{P}$ in the
following way:
\begin{equation}
\e R_1=R\sin{u},\quad \g R_2 = R\cos{u},\quad \Omega = zR,\quad v =
\theta_2-2\theta_1+2\beta_1-\beta_2\;.
\label{eq:params}
\end{equation}
Now equation~(\ref{eq:bkm_th1}) takes the form:
\begin{equation}
\dot{\psi} =R\left(x - z -\sin{u}\sin{\psi}-\cos{u} \sin(2\psi-v)\right) =
R\left(x - z - y(u,v,\psi)\right)\;.
\label{eq:bkm_th2}
\end{equation}
Here we denoted $x = \omega/R$ and $y(u,v,\psi)=\sin u \sin \psi + \cos
u\sin(2\psi-v)$.
 
Setting parameters $\mathbf{P}$ to some constant values in (\ref{eq:bkm_th2})
[this means that $R_{1,2},\theta_{1,2}$ are constants, i.e. 
the order parameters are uniformly rotating with velocity $\Omega$],
one can find a stationary distribution function $\rho(\psi|x,\mathbf{P})$ 
and then
calculate corresponding complex order parameters as:
\begin{equation}
\begin{aligned}
&R_1e^{i\theta_1} = e^{i(\theta_1-\beta_1)}R\iint dxd\psi\rho(\psi|x,\mathbf{P})
e^{i\psi}g(Rx) = e^{i(\theta_1-\beta_1)}R F_1(\mathbf{P}) e^{iQ_1(\mathbf{P})}\\
&R_2e^{i\theta_2} = e^{i2(\theta_1-\beta_1)}R\iint
dxd\psi\rho(\psi|x,\mathbf{P}) e^{i2\psi}g(Rx) = e^{i2(\theta_1-\beta_1)}R
F_2(\mathbf{P}) e^{iQ_2(\mathbf{P})}\\
&F_{m}(\mathbf{P}) e^{iQ_{m}(\mathbf{P})}\equiv\iint 
dxd\psi\rho(\psi|x,\mathbf{P})
e^{im\psi}g(Rx)\qquad
m=1,2\;.
\end{aligned}
\label{eq:int}
\end{equation}

\begin{figure}[!htb]
\centerline{(a)\includegraphics[width=0.49\columnwidth]{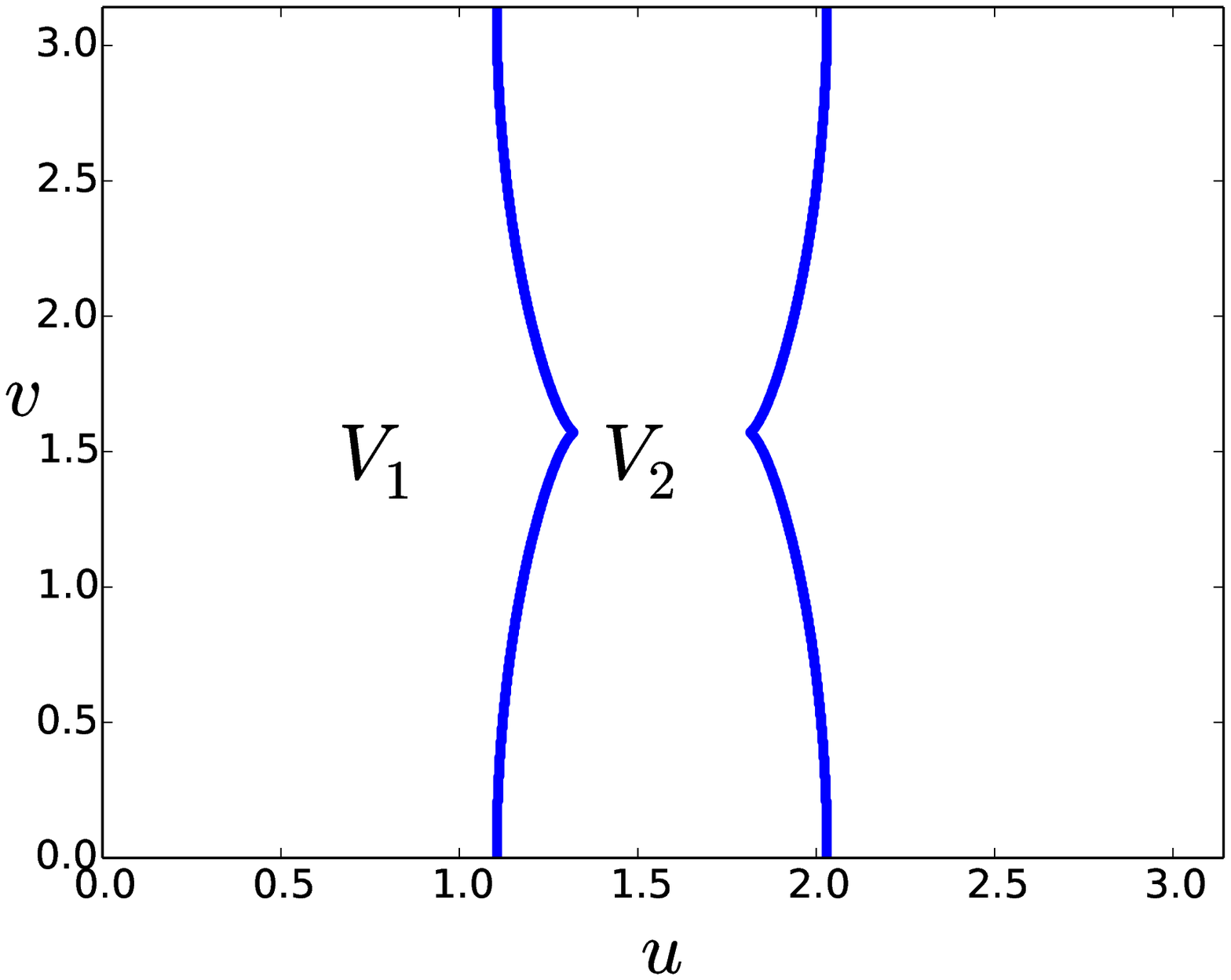}
(b)\includegraphics[width=0.49\columnwidth]{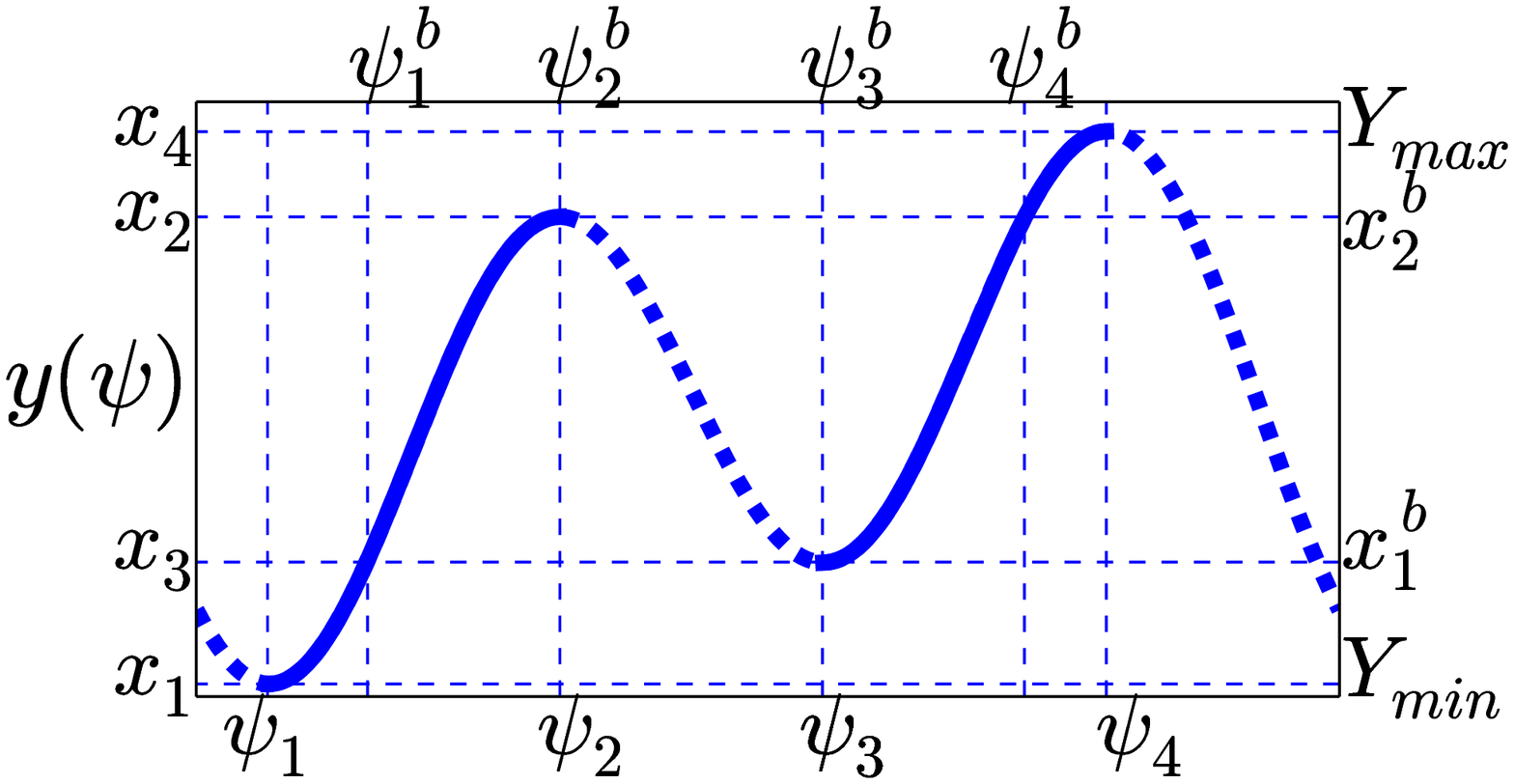} }
\centerline{(c)\includegraphics[width=0.49\columnwidth]{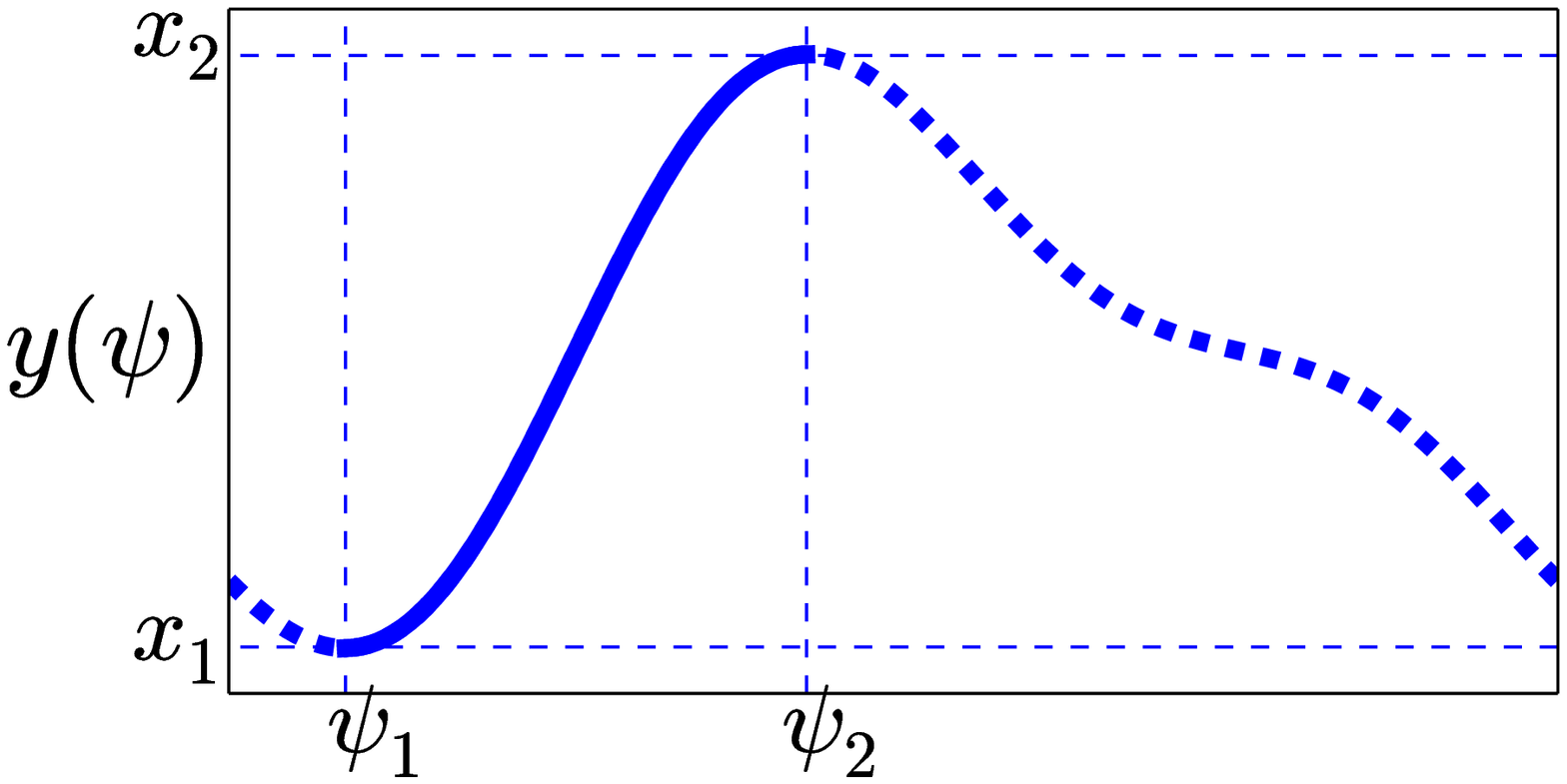}
(d)\includegraphics[width=0.49\columnwidth]{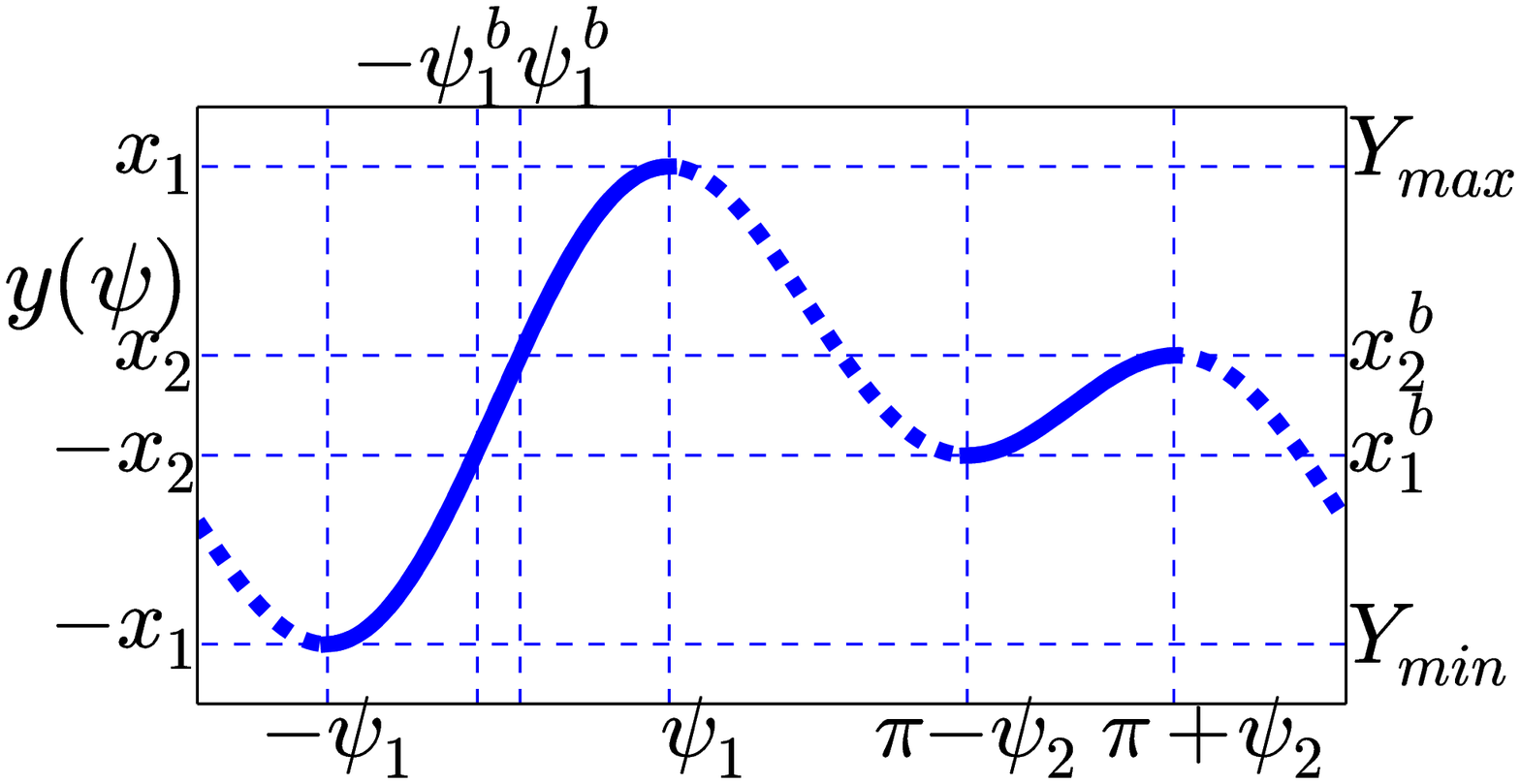} }
\caption{(a) Regions $V_1$ and $V_2$ in the plane of
parameters $(u,v)$: 
Domain $V_1$ corresponds to a double-well form of function $y(u,v,\psi)$
(Fig.~\ref{fig:locked_ph}(b,d)), while in $V_2$ 
$y(u,v,\psi)$ has a single-well form
like sown in Fig.~\ref{fig:locked_ph}(c). 
(b) Example of function $y(u,v,\psi)$ with 4 extrema is presented. There are two
stable branches (solid curves) for stationary phases of locked oscillators. The
 left branch $\psi=\Psi_1(x,\mathbf{P})$ is larger than the right one
$\psi=\Psi_2(x,\mathbf{P})$. $(\psi_{1,2},x_{1,2})$ denote coordinates of the
extrema corresponding to the  branch $\Psi_1$, while
$(\psi_{3,4},x_{3,4})$ denotes extrema at $\Psi_2$. 
(c) Example of function $y(u,v,\psi)$ with only two extrema and one stable
branch $\psi=\Psi_1(x,\mathbf{P})$  (solid curve). 
(d) Example of function $y(u,v,\psi)$ in the special case $v=0$.}
\label{fig:locked_ph}
\end{figure}

Our next goal is to calculate the integrals $F_m(\mathbf{P})$, for this we need 
to find, using the dynamical equation (\ref{eq:bkm_th2}), 
the distribution function $\rho(\psi|x,\mathbf{P})$.
Let $Y_{min}$ and $Y_{max}$ denote the global minimum and the 
global maximum of function
$y(u,v,\psi)$, correspondingly (Fig.\ref{fig:locked_ph}(b)). 
All the oscillators can be separated into locked ones (for
$Y_{max}\geq|x-z|\geq Y_{min}$) or rotating, unlocked 
ones ($x-z>Y_{max}$ or $x-z<Y_{min}$). 
The distribution function of rotating oscillators (index $r$) is inversely 
proportional to
their phase velocity:
\begin{equation}
\rho_r(\psi|x,\mathbf{P}) = g(Rx)\rho(\psi|x,\mathbf{P})= 
\frac{C(x)}{|x-z-y(\psi,u,v)|}\;,
\label{eq:rdist}
\end{equation}
where $C(x)$ is the normalization constant to which we included also the distribution
of frequencies:
$$
C(x) = \frac{g(Rx)}{\int_0^{2\pi}\frac{d\psi}{|x-z-y|}}.
$$

The stationary phases of locked oscillators (index $l$)
can be found from the following
relation:
\begin{equation}
x - z = y(u,v,\psi)\;.
\label{eq:locked_ph}
\end{equation}
When finding $\psi$ as a function of $x$, we have 
to satisfy an additional stability
condition $\frac{\partial y(u,v,\psi)}{\partial \psi}>0$ that follows from
the dynamical equation
(\ref{eq:bkm_th2}).
In the $(u,v)$ plane there are two regions $V_1$ and $V_2$
(Fig.~\ref{fig:locked_ph}(a)) which produce qualitatively different properties
of system (\ref{eq:bkm_th2}) and different types of distribution function
$\rho_l(\psi|x,\mathbf{P})$:

\paragraph{(i) $\{u,v\}\in V_1$} In this case function $y(u,v,\psi)$ has a
double-well form like shown in Fig.\ref{fig:locked_ph}(b). 
According to (\ref{eq:locked_ph}), oscillators can be located on two possible
stable branches highlighted by solid curves in Fig.\ref{fig:locked_ph}(b): the
first branch is $\psi = \Psi_1(x,\mathbf{P})$ in the range $\psi\in
[\psi_1,\psi_2]$ and another branch is $\psi = \Psi_2(x,\mathbf{P})$ for
$\psi\in [\psi_3,\psi_4]$.
Here and below we assume $\Psi_1(x,\mathbf{P})$ to be the biggest stable branch.
In the range $(x-z)\in (x^b_1,x^b_2)$ (Fig.~\ref{fig:locked_ph}(b)) there is an
area of bistability on the microscopic level: the oscillators with the 
same natural
frequency $x$ can be locked at two 
different phases $\Psi_1(x,\mathbf{P})$ and
$\Psi_2(x,\mathbf{P})$.
Therefore, the distribution function has the following form:
\begin{equation}
\rho_{l}(\psi|x,\mathbf{P})=\left\{
\begin{array}{l}
(1-S(x))\delta(\psi-\Psi_1(x,\mathbf{P}))+S(x)\delta(\psi-\Psi_2(x,\mathbf{P}
))\\ 
\text{for $(x-z)\in(x^b_1,x^b_2)$}\\
\delta(\psi-\Psi_1(x,\mathbf{P}))\ \ \ \text{for $(x-z)\in [x_1,x_2]\setminus
(x_1^b,x_2^b)$}\\
\delta(\psi-\Psi_2(x,\mathbf{P}))\ \ \ \text{for $(x-z)\in [x_3,x_4]\setminus
(x_1^b,x_2^b)$}\\
\end{array}
\right.
\label{eq:distr_func_1}
\end{equation}
Here $0\leq S(x)\leq 1$ is an indicator function describing the 
redistribution over the
stable brunches; this function is arbitrary.

\paragraph{(ii) $\{u,v\}\in V_2$} In the second case, 
 function $y(u,v,\psi)$ has
only two extrema (Fig.~\ref{fig:locked_ph}(c)) and there is only one stable
branch $\psi = \Psi_1(x,\mathbf{P})$. The distribution function is:
\begin{equation}
\rho_{l}(\psi|x,\mathbf{P})=\delta(\psi-\Psi_1(x,\mathbf{P}))\ \text{for $x\in
(z+x_1,z+x_2)$}
\label{eq:distr_func_2}
\end{equation}

Taking into account the obtained expressions for the distribution function
(\ref{eq:rdist},\ref{eq:distr_func_1},\ref{eq:distr_func_2}), 
the integrals in (\ref{eq:int})
can be rewritten as follows:
\begin{equation}
\begin{aligned}
&F_m(\mathbf{P})e^{iQ_m(\mathbf{P})} = \int_{\psi_1}^{\psi_2}d\psi e^{im\psi}
g\left(R(z+y)\right) \frac{\partial y}{\partial \psi}-\\
&\int_{\psi^b_1}^{\psi^b_2}d\psi e^{im\psi}S(z+y) g\left(R(z+y)\right)
\frac{\partial y}{\partial \psi}+
\int_{\psi_3}^{\psi_4}d\psi e^{im\psi} g\left(R(z+y)\right) \frac{\partial
y}{\partial \psi}-\\
&\int_{\psi^b_3}^{\psi^b_4}d\psi e^{im\psi}\left(1-S(z+y)\right)
g\left(R(z+y)\right) \frac{\partial y}{\partial \psi}
+\int_{\mathcal{X}}\int_{0}^{2\pi}dxd\psi \frac{C(x)e^{im\psi}}{|x-z-y|}
\end{aligned}
\label{eq:int1}
\end{equation}
Here in the last integral we denote the interval $\mathcal{X} =
(-\infty,z+Y_{min}) \bigcup (z+Y_{max},\infty)$.

Now, using the 
integrals (\ref{eq:int1}), one can calculate the absolute values of the
complex
order parameters $R_{1,2}$ and the frequency $\Omega$ as functions of
introduced parameters
$R$, $u$, $v$, $z$:
\begin{equation}
R_{1,2}(\mathbf{P}) = RF_{1,2}(\mathbf{P}),\ \Omega(\mathbf{P}) = Rz\;.
\label{eq:self_cons1}
\end{equation}
Then, from relations (\ref{eq:params}), (\ref{eq:int}) and
(\ref{eq:self_cons1}) it follows that:
\begin{equation}
\e(\mathbf{P})=\frac{\sin{u}}{F_1(\mathbf{P})},\
\g(\mathbf{P})=\frac{\cos{u}}{F_2(\mathbf{P})},\ \beta_1(\mathbf{P}) =
Q_1(\mathbf{P}),\ \beta_2(\mathbf{P}) = Q_2(\mathbf{P})-v\;.
\label{eq:self_cons2}
\end{equation}
All together equations (\ref{eq:self_cons1}) and (\ref{eq:self_cons2}) determine
the stationary amplitudes of the order parameters $R_{1,2}$ and
 the frequency of their rotation $\Omega$ in dependence on model parameters
$\varepsilon$, $\gamma$, $\beta_{1,2}$ in an analytic, albeit parametric form.
Note that this solution fully accounts to multi-branch entrainment, due to 
presence of the indicator function $S$. 
Arbitrariness of this functions means that
there is a huge multiplicity of microstates.

We stress, that in the solution (\ref{eq:self_cons1},\ref{eq:self_cons2})
parameters $R$, $u$, $v$, $z$ and the indicator function 
are \textit{independent}, while
the order parameters $R_{1,2}$ and the coupling parameters 
$\e,\gamma,\beta_{1,2}$ are functions of them.
If, on the other hand, one wants to fix the coupling parameters, 
then one should adjust some of the parameters  $R$, $u$, $v$, $z$ and 
the indicator function, which will be now not independent. This is 
a standard procedure in a parametric representation of a solution.

\section{Symmetric bi-harmonic coupling function}
\label{sec:sym}
Here we consider the simplest case where $\beta_1=\beta_2=0$, what 
corresponds to 
a symmetric
coupling function $\Gamma(x)=\varepsilon\sin(x)+\gamma\sin(2x)$.

\subsection{General solution of self-consistent equations}
Due to the symmetry of the coupling function, it is possible to  perform the
 self-consistent approach in the special case
$z=v=0$ (see however Section~\ref{sec:nonsym} for a more general situation). 
First we will simplify equations
(\ref{eq:int1},\ref{eq:self_cons1},\ref{eq:self_cons2}) taking into account the
relation $z=v=0$.

A typical form of function $y(u,v=0,\psi)$ is presented in
Fig.~\ref{fig:locked_ph}(d). 
For $v=0$, the critical value $u = \pm \arctan(2)$  
separates double-well and single-well shapes
of function $y(u,0,\psi)$. If $|\tan(u)|<2$, the 
function $y(u,0,\psi)$ contains two 
stable branches $\Psi_1$ and $\Psi_2$ (see Fig.~\ref{fig:locked_ph}(d)), 
otherwise only one branch $\Psi_1$ exists like it is 
shown in Fig.~\ref{fig:locked_ph}(c).
The stable branches $\Psi_{1}$
and $\Psi_2$ \textbf{(if exists)} are always centered in the intervals 
$$\Psi_1:\ [-\psi_1,+\psi_2]\ \text{and}\ \Psi_2:[\pi-\psi_2,\pi+\psi_2],$$ 
where the values $\psi_{1,2}$ can be calculated explicitly:
$$
\psi_{1,2}=\arccos\left(\frac{\mp \sin u+
\sqrt{\sin^2u+32\cos^2 u}}{8\cos u}\right)
$$
Moreover, the branches $\Psi_{1,2}$ are symmetric (see
Fig.\ref{fig:locked_ph}(d)):
$$
y(u,0,\psi) = -y(u,0,-\psi),\ y(u,0,\pi+\psi) = -y(u,0,\pi-\psi).
$$
 Taking all this into account, the relations
(\ref{eq:int1}) can be radically simplified:
\begin{equation}
\begin{aligned}
&F_m(R,u)e^{iQ_m(R,u)} = \int_{-\psi_1}^{\psi_1}d\psi e^{im\psi}
S(y)g\left(R(y)\right) \frac{\partial y}{\partial \psi}+\\
&\int_{\pi-\psi_2}^{\pi+\psi_2}d\psi e^{im\psi} (1-S(y))g\left(R(y)\right)
\frac{\partial y}{\partial \psi}+
\int_{|x|>x_1}\int_{0}^{2\pi}dxd\psi \frac{C(x)e^{im\psi}}{|x-z-y|}\;.
\end{aligned}
\label{eq:int2}
\end{equation}
Here we assumed that $S(y)=1$ everywhere outside interval $[-\psi_2,\psi_2]$
(see Fig.~\ref{fig:locked_ph}).
If the functions $S(x)$ and $g(x)$ are even, then it is easy to see that the
imaginary part in all of the integrals in (\ref{eq:int2}) vanishes 
(recall that
$y(u,0,\psi)$ is odd).
Thus, for any $S(x)=S(-x)$ and $g(x)=g(-x)$ we obtain $Q_{1,2}(R,u) = 0$ and
automatically $\beta_{1,2} = 0$. (See Section~\ref{sec:nonsym} 
below for discussion of an asymmetric indicator function $S$.)

In summary, for the 
case $z=v=0$ and even $S(x)$, $g(x)$  we have $\Omega = \beta_{1,2}=0$
and the following expressions for the parameters $\e,\gamma$ and real order 
parameters $R_{1,2}$
as functions of two introduced parameters $R,u$:
\begin{equation}
R_{1,2}(R,u) = RF_{1,2}(R,u),\ \e(R,u)=\frac{\sin{u}}{F_1(R,u)},\
\g(R,u)=\frac{\cos{u}}{F_2(R,u)}.
\label{eq:self_cons_vz}
\end{equation}

\subsection{Stability of the incoherent state}
\label{sec:stis}
\begin{figure}
\centerline{(a)\includegraphics[width=0.5\columnwidth]{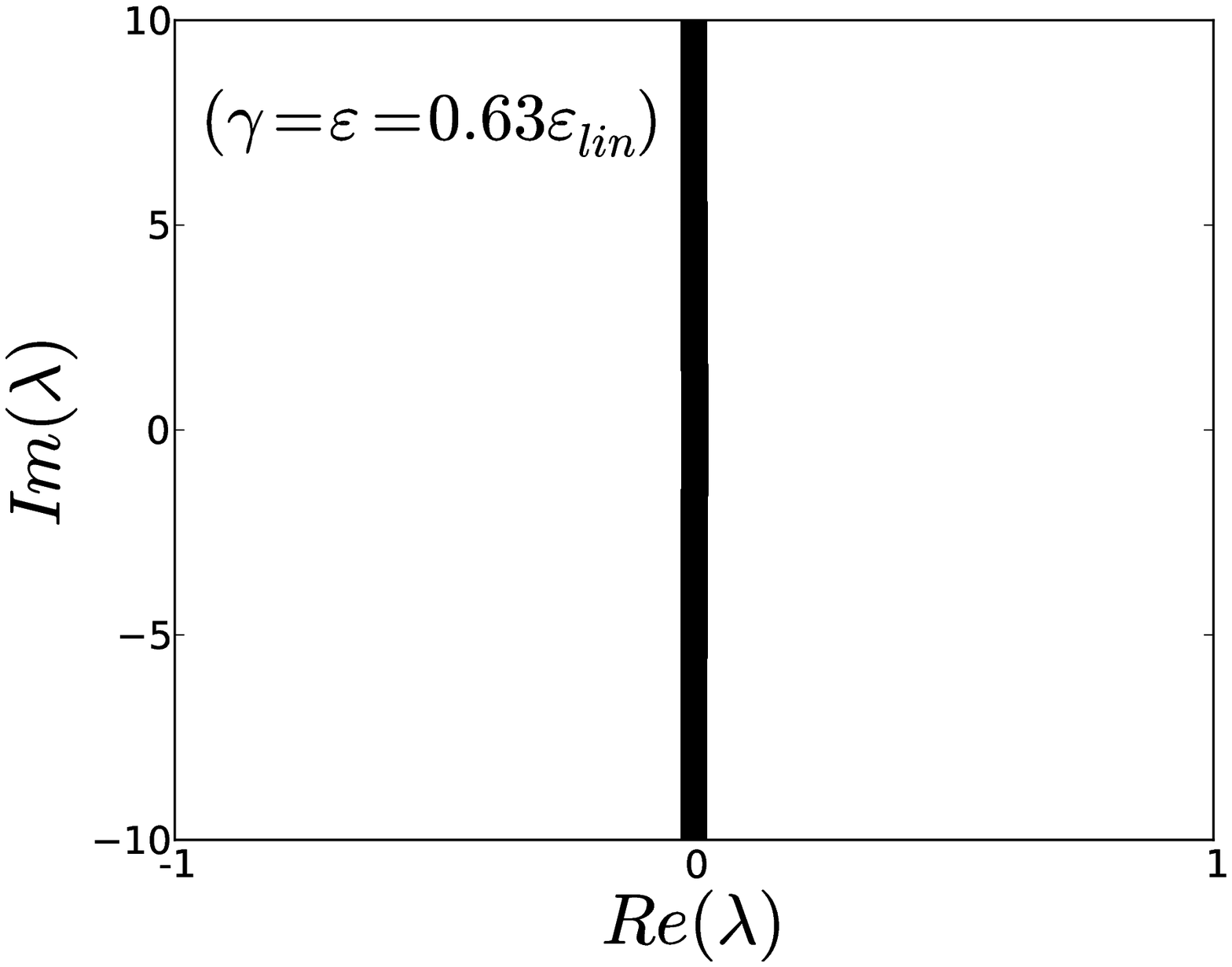}
(b)\includegraphics[width=0.5\columnwidth]{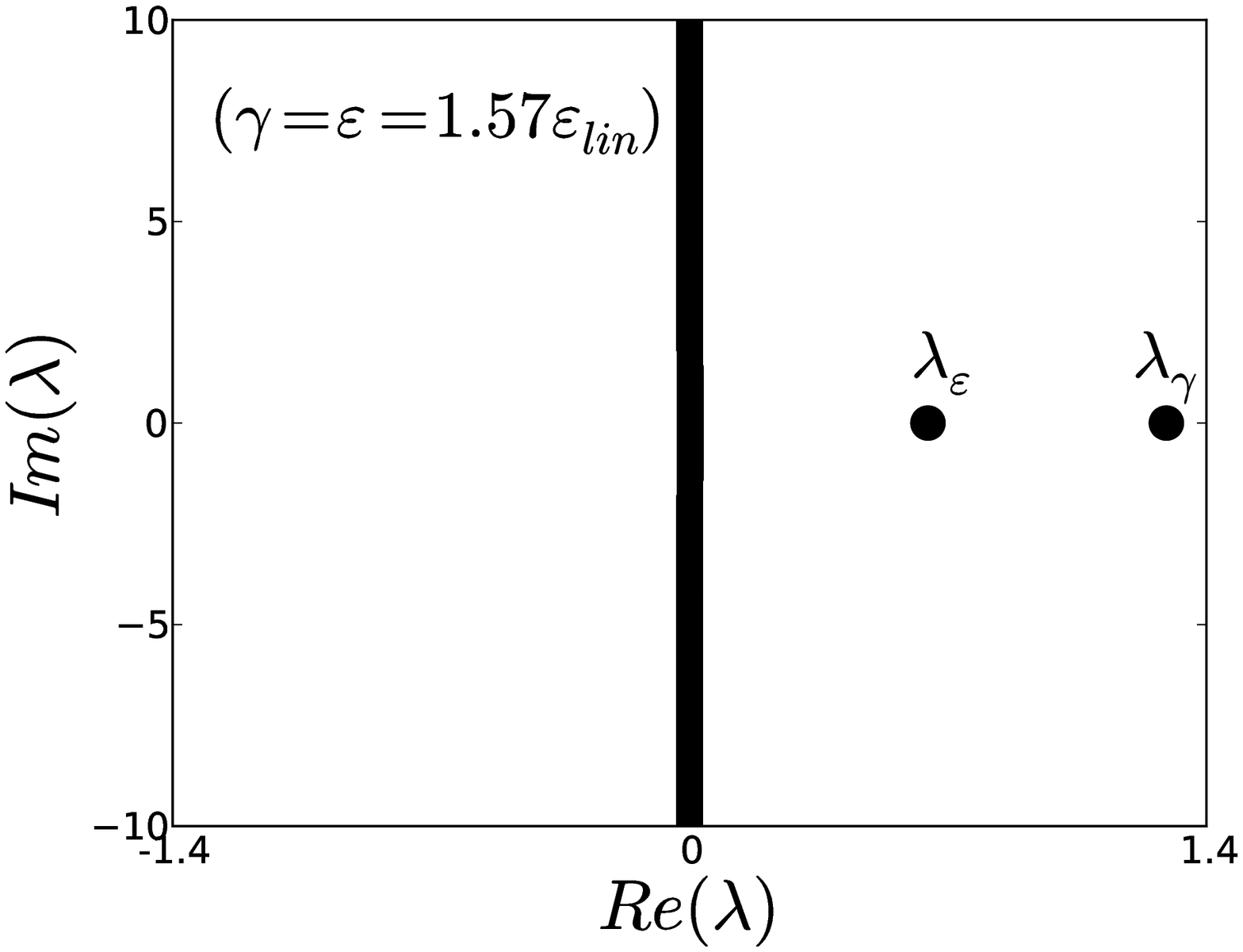} }
\caption{Illustration of the spectra of incoherent state in two different
regions: (a) continuous part of the spectrum lies on the imaginary axes revealing
neutral stability in the region $\gamma<\gamma_{lin}$, $\e<\e_{lin}$,
(b) when on of the couplings exceeds threshold $\e>\e_{lin}$ or
$\gamma>\gamma_{lin}$ the eigenvalue with positive real part appears in the
discrete part of the spectrum.
Both calculation were made for the Gaussian distribution of frequencies
$g(\w)=\frac{1}{2\pi}e^{-\w^2/2}$,
$\e_{lin}=\gamma_{lin}=2\sqrt{\frac{2}{\pi}}$.
}
\label{fig:stab}
\end{figure}

Before proceeding with presentation of the main results we recall that an
issue of linear stability of the incoherent state (with uniform distribution of phases)
 was a milestone in almost all
preceding mathematical
studies~\citep{Strogatz-Mirollo-91,Crawford-95,Crawford-Davies-99,
Chiba-Nishikawa-11} of Kuramoto-type models.
This analysis of the partial differential equation 
for the density distribution function
revealed the following stability properties of the incoherent
state~\citep{Strogatz-Mirollo-91,Crawford-95,Crawford-Davies-99,
Chiba-Nishikawa-11}: 
(i) the continuous part of the spectrum always lies on the imaginary axis;
(ii) when
one of the couplings exceeds certain threshold $\e>\e_{lin}$ or
$\gamma>\gamma_{lin}$,
in the discrete spectrum appears an eigenvalue ($\lambda_\e$ or $\lambda_\gamma$
correspondingly) with a positive real part revealing
instability of the asynchronous state. 
We illustrate this in Fig.~\ref{fig:stab}.
 In the linear theory, the modes of the perturbation corresponding to
the harmonics of the coupling are independent on each other, and one gets 
$\e_{lin}=\gamma_{lin}=\frac{2}{\pi g(0)}$. Below in this paper we use
a  Gaussian
distribution of frequencies 
$g(\w)=(2\pi)^{-1/2}\exp(-\w^2/2)$, thus 
$\e_{lin}=\gamma_{lin}=2\sqrt{\frac{2}{\pi}}$. In
presentation of the results, we will 
always
normalize the values of 
the coupling parameters
$\e,\gamma$ by the linear stability thresholds.

\subsection{Diagram of synchronous states}

\begin{figure}
\centerline{(a)\includegraphics[width=0.5\columnwidth]{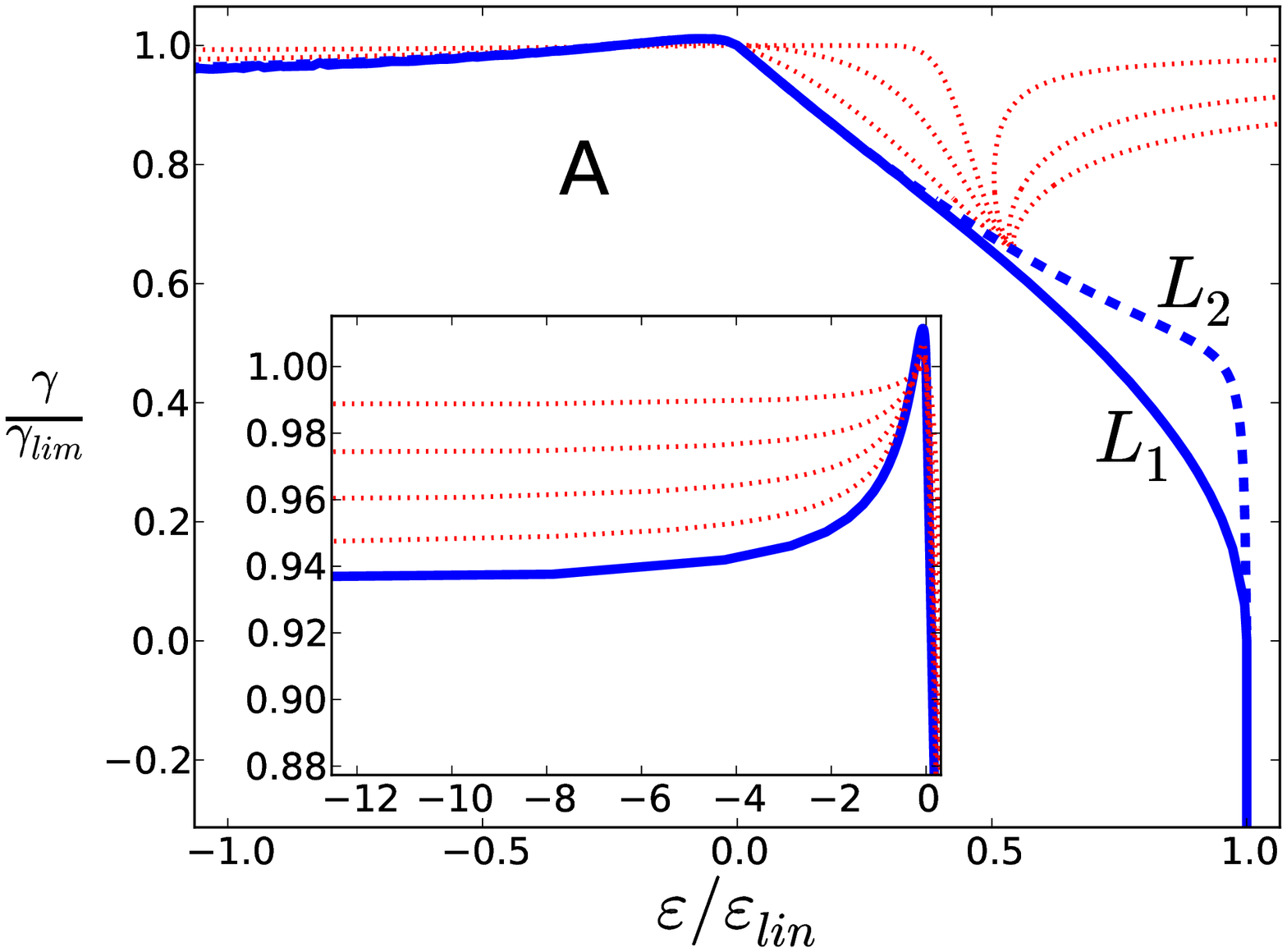}
(b)\includegraphics[width=0.5\columnwidth]{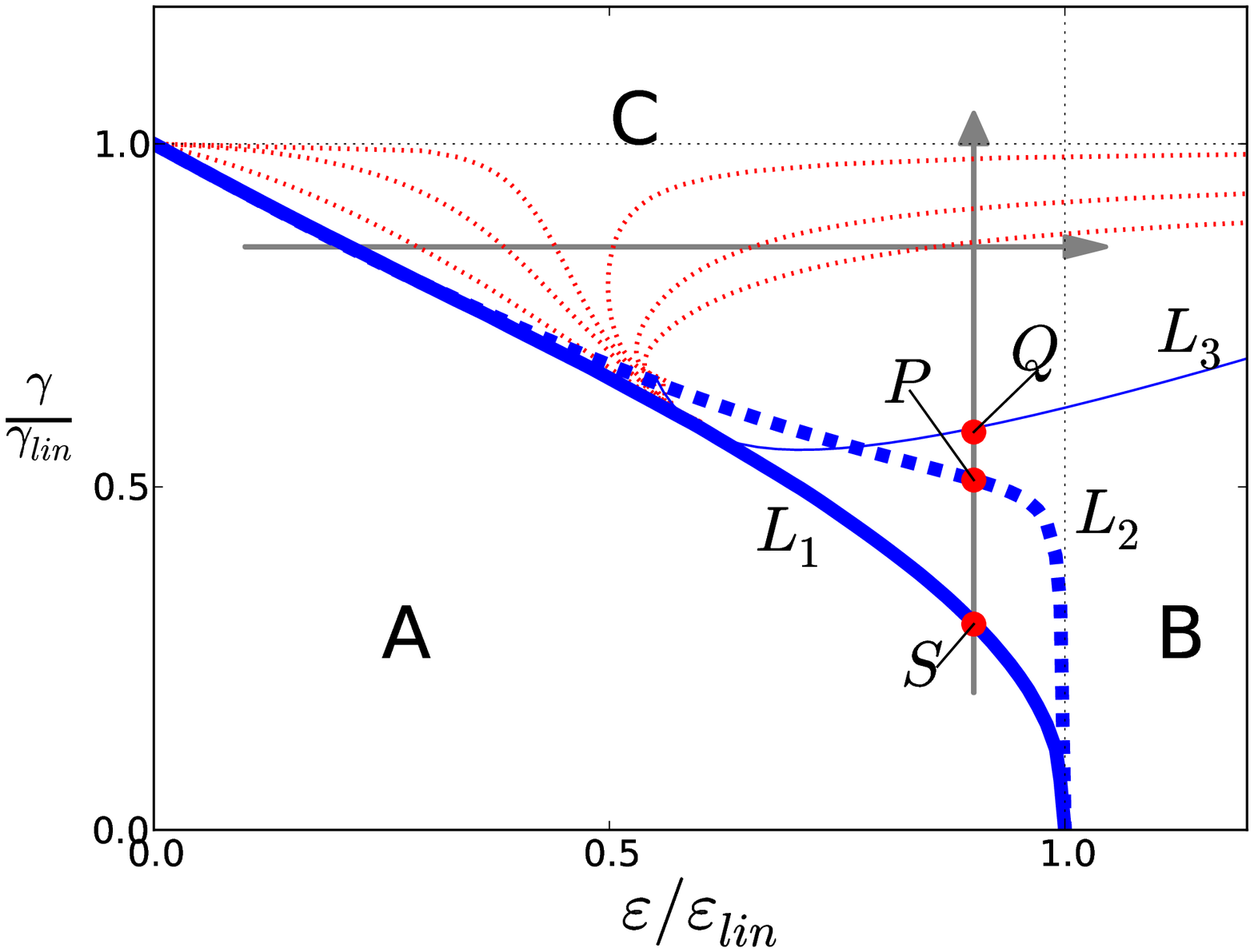} }
\caption{(a) Diagram of
different synchronous states in dependence
on 
parameters $(\e,\gamma)$, resulting from the analytical solution
Eqs.~(\ref{eq:int2},\ref{eq:self_cons_vz}). Bold (blue) line $L_1$: border of
synchronous states, inside area A there is only the incoherent solution;
bold
dashed (blue) line $L_2$: order parameters vanish. Between lines $L_1$ and $L_2$
there are two  
solutions (stable and unstable) with non-zero $R_{1,2}$ and the transition to
synchrony is hard
(see region between points $S$ and $P$ in Fig.~\ref{fig:gam_08}(a)).
Dotted (red) lines: onset of
synchrony for $\sigma=0.2,\,0.4,\,0.5,\,0.6,\,0.8,\,1$ (from left to right).
Inset shows the domain $\e<0$ in more details (with the same axes).
(b) The same as in Fig.~\ref{fig:planeepsgam}(a) but in the area $\e,\
\gamma>0$. An additional line $L_3$ is drawn from the condition 
$\tan u = 2$, dividing domains B (single synchronous state)
and C (multiple synchronous
states). Above
$L_3$ multiplicity of
synchronous states due to multi-branch entrainment occurs 
(beyond point $Q$ in Fig.~\ref{fig:gam_08}(a)).}
\label{fig:planeepsgam}
\end{figure}

\begin{figure}
\centerline{(a)\includegraphics[width=0.49\columnwidth]{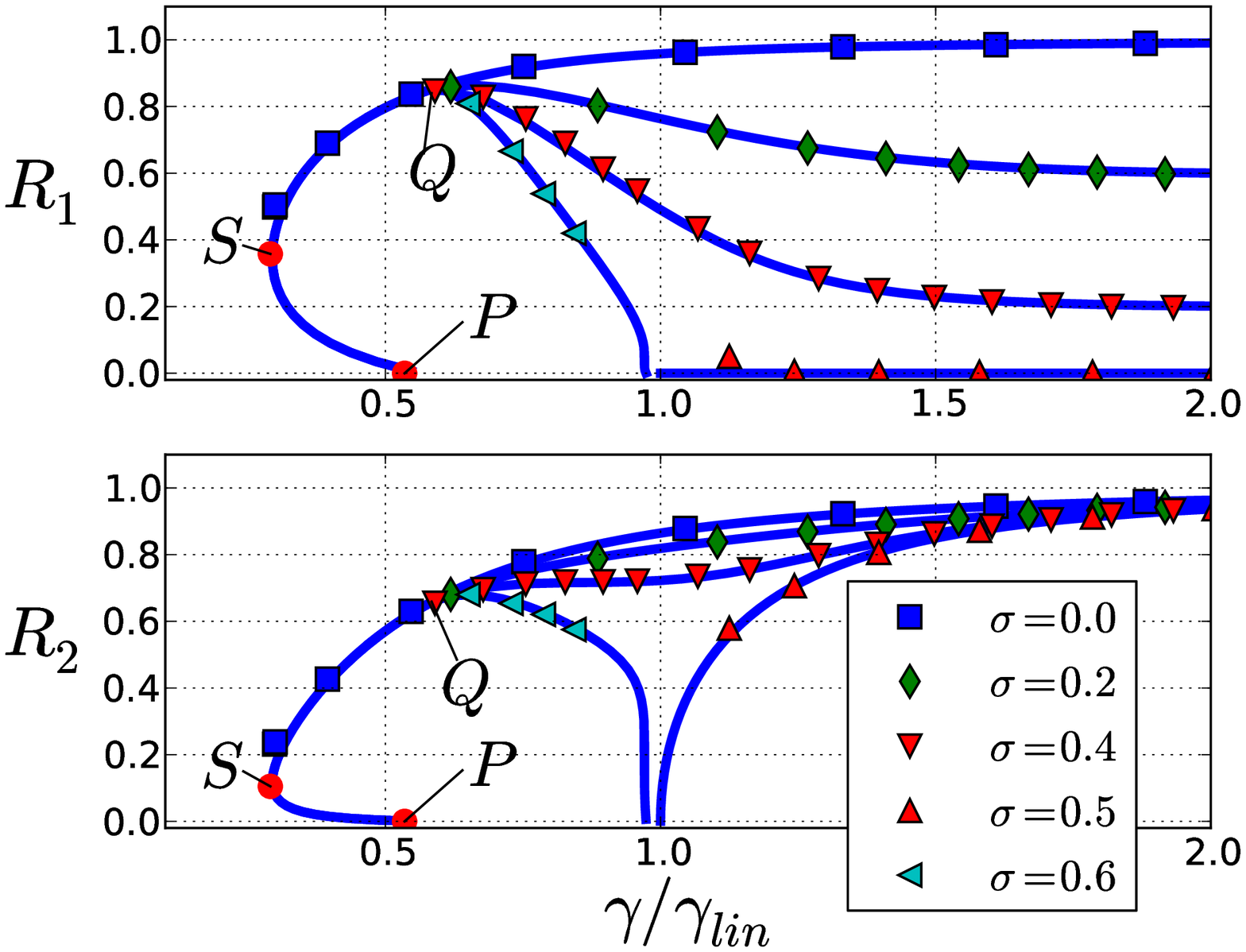}
(b)\includegraphics[width=0.49\columnwidth]{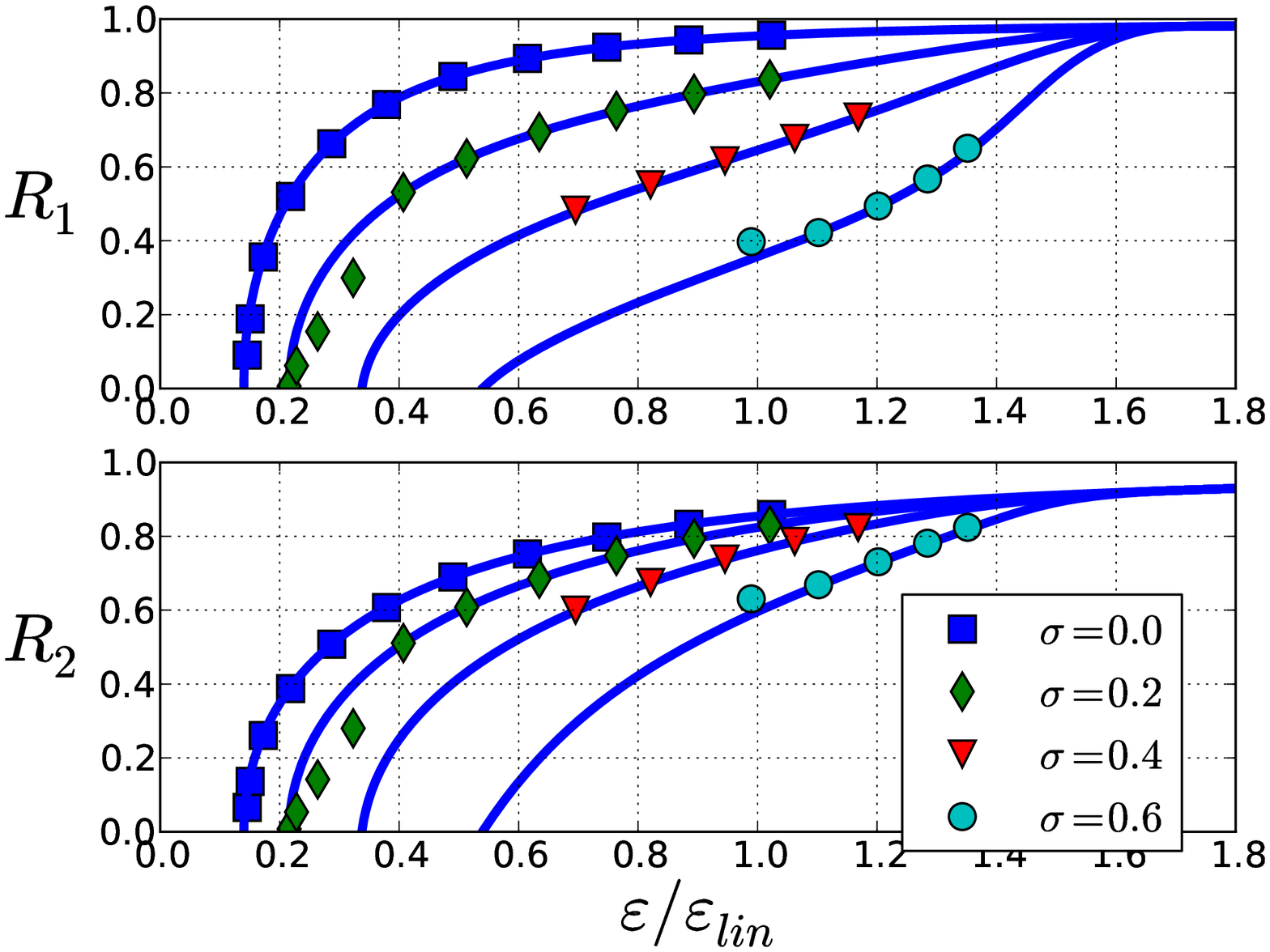} }
\centerline{(c)\includegraphics[width=0.49\columnwidth]{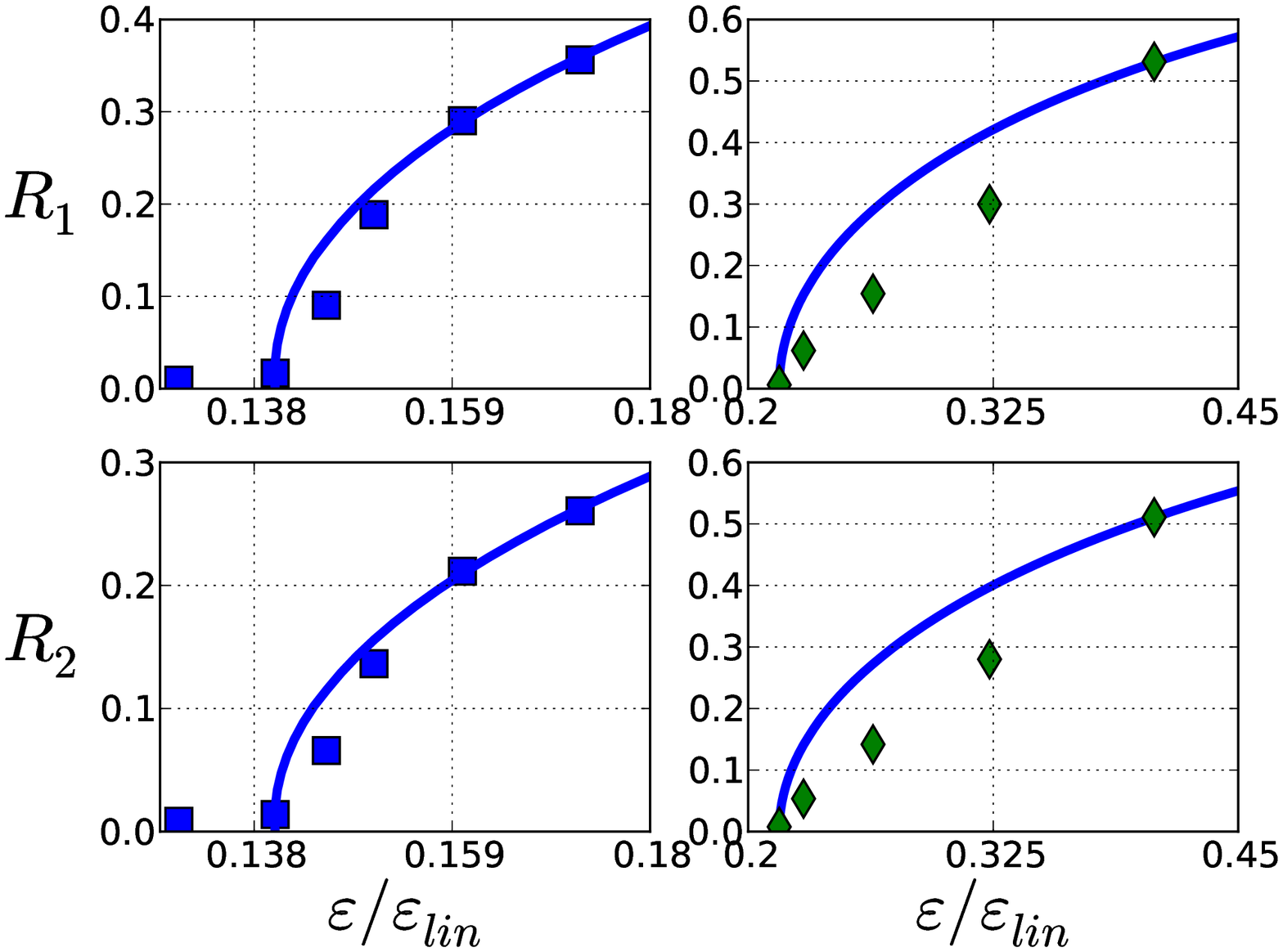}
(d)\includegraphics[width=0.49\columnwidth]{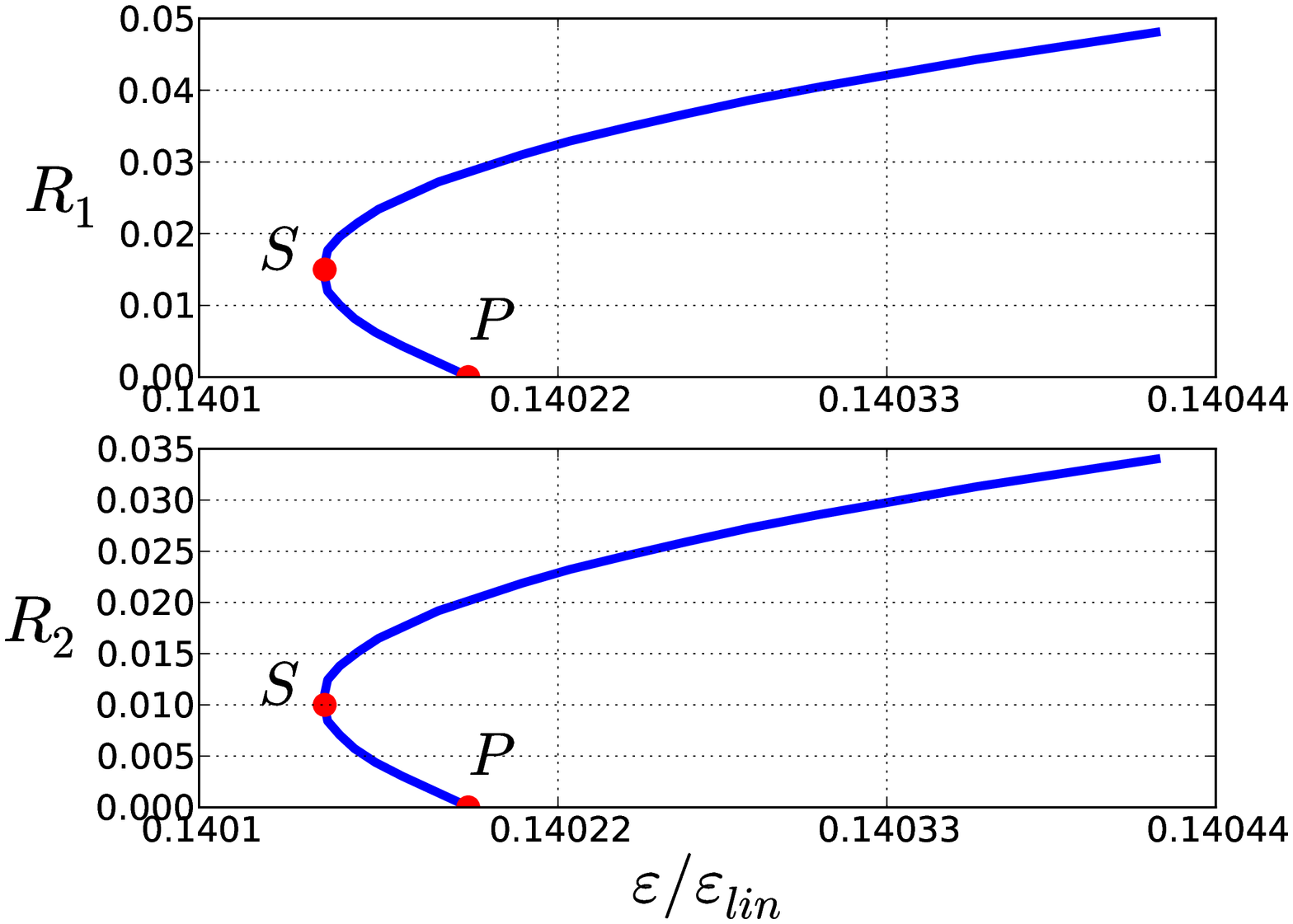} }
\caption{(a) Dependence of the order parameters
$R_{1,2}$ on coupling strength $\gamma$ at fixed value of $\e=0.9\e_{lin}$ (see
also vertical arrow in Fig.~\ref{fig:planeepsgam}(b)).  Markers are results of
direct simulation
of a population of $N=2\cdot 10^{4}$ oscillators.
Different curves correspond to different
values of $\sigma$, as depicted on the panel. For
 $\gamma\lesssim  0.6\gamma_{\text{lin}}$ there is a unique synchrony state,
for larger couplings
multiplicity is observed.
Point $S$ denotes a ``saddle-node bifurcation'' at which coherent states appear
(curve $L_1$ in Fig.~\ref{fig:planeepsgam}).
At point $P$ the order parameters at the
unstable branch of coherent solution vanish (curve $L_2$ in
Fig.~\ref{fig:planeepsgam}).
 Between points $S$ and $P$ a finite perturbation of the incoherent state is
needed to come to 
a synchronous regime.
Point $Q$, the onset of multiplicity, corresponds to curve $L_3$ in
Fig.~\ref{fig:planeepsgam}(b). (b) The same as in Fig.~\ref{fig:gam_08}(a) but
for $\gamma=0.9\gamma_{lin}$, and varying $\e$. 
For $\e\gtrsim  1.6\e_{\text{lin}}$ the solution is unique, for  smaller $\e$
there are multiple states with different $\sigma$
appearing at different critical couplings.
 (c) Detailed view of curves $R_{1,2}(\e)$ for $\sigma=0$ (left
panels) and $\sigma=0.2$ (right panels). For both cases
$\gamma=0.9\gamma_{lin}$. 
Here markers denote averaged values of stationary order parameters of different
independent numerical simulations (see text). 
(d) Enlargement of the curves $R_{1,2}(\e)$ ($\sigma=0$,
$\gamma=0.9\gamma_{lin}$, panel(b)) for small values of order parameters, 
indicating a  first-order type of the transition hardly seen in (b).
}
\label{fig:gam_08}
\end{figure}

In Figs.~\ref{fig:planeepsgam} we illustrate the diagram of the states
on the plane of parameters $(\e,\gamma)$, and in Fig.~\ref{fig:gam_08} 
some cuts of it,
 for the
simplest case, where the indicator function $S(\w)=\sigma$ is a constant.
This diagram is obtained by application of analytic formulas 
(\ref{eq:self_cons_vz}).

We start the description with an even simpler case 
$\sigma=0$ (so that all the phases are on one stable
branch).
Setting 
in \eqref{eq:int2},\eqref{eq:self_cons_vz}
$R=0^{+}$ and varying $u$, we find a curve on the plane of parameters
$(\e,\gamma)$ where
the order parameters $R_{1,2}$ vanish (line $L_2$ in
Fig.~\ref{fig:planeepsgam}, see Section~\ref{sec:perturb} below for 
the details of calculation of this line).  
Remarkably, solutions $R_{1,2}(\e,\gamma)$ behavior characteristic
for first-order phase
transitions, as the coupling 
strengths $(\e,\gamma)$ increase (Fig.~\ref{fig:gam_08}a; exception
are the pure cases $\e=0$ and $\gamma=0$, see Section~\ref{sec:perturb} below). 
Therefore, in the plane $(\e,\gamma)$ also exists the curve $L_1$ which
corresponds to the line of a ``saddle-node bifurcation'' where two branches of
coherent solutions first appear (point $S$ in Fig.~\ref{fig:gam_08}a).
This line $L_1$ split the plane $(\e,\gamma)$ in two different regions: in
area $A$ in Fig.~\ref{fig:planeepsgam}(a,b) only incoherent solution of self-consistent equations 
exists, outside
area $A$ (regions $B$ and $C$ in Fig.~\ref{fig:planeepsgam}(b)) synchronous
solution(s) exist.
Between curves $L_1$ and $L_2$ there are two solutions with $\sigma=0$. 
We also show a curve $L_3$ corresponding to the parameter value 
$\tan u=2$, which
separates the two-branch
(Fig.~\ref{fig:locked_ph}(a,d)) and
the one-branch (Fig.~\ref{fig:locked_ph}(b)) situations (marked as C and B
on panel Fig.~\ref{fig:planeepsgam}(b)
correspondingly). 

Below $L_3$ there is a solution with $S(\w)=0$ only,
above it, 
multiplicity due to arbitrariness of the
indicator function $S(\w)$ occurs. We
depict also  
curves corresponding  to synchronous solutions
with $R_{1,2}=0^{+}$ at several fixed values of $\sigma$ (red curves in 
Fig.~\ref{fig:planeepsgam}), 
to
the right of these curves 
synchronous states with corresponding values of $\sigma$ exist.

We illustrate different synchronous regimes as functions of coupling parameters 
$(\e,\gamma)$ in Figs.~\ref{fig:gam_08}(a,b).
Fig.~\ref{fig:gam_08}(a) shows dependence of synchronous states on the coupling
parameter $\gamma$ for fixed $\e = 0.9\e_{\text{lin}}$ (vertical arrow in
Fig.~\ref{fig:planeepsgam}(b)).
As it has been mentioned above, two branches of 
coherent solutions arise at point
$S$. 
With increase of $\gamma$, the lower branch merge with incoherent solution at
point $P$. 
The upper branch is unique until
the border of multiplicity $\tan u=2$ (point $Q$) is crossed.
Multiple solutions exist
for all larger values
of $\gamma$. 

A special symmetric solution appears at the linear threshold $\gamma=\gamma_{lin}$.
This regime contains only the second harmonic ($R_1=0$) and has symmetric 
redistribution of oscillators ($\sigma=0.5$)
between the two symmetric stable branches.
This regime appears as a 
square root of supercriticality $R_2\sim (\gamma-\gamma_c)^{\frac{1}{2}}$ (see 
the branch of $R_2$ starting at $\gamma/\gamma_{lin}=1$ for $\sigma=0.5$
in Fig.~\ref{fig:gam_08}(a)) and  
corresponds to the bifurcation from the
asynchronous state
as described in~\cite{Crawford-95,Crawford-Davies-99}.

In Fig.~\ref{fig:gam_08}(b) the order parameters are shown as functions of $\e$ 
for fixed
$\gamma = 0.9\gamma_{lin}$(horizontal arrow in
Fig.~\ref{fig:planeepsgam}(b)). As here almost everywhere we are in the region
of multiplicity,  the synchrony arises at different values
of $\e$ for different $\sigma$, 
and immediately beyond the threshold (which corresponds to $\sigma=0$) 
multiple synchrony states with $\sigma>0$
are possible (as here $\tan u <2$).  
With further increase of $\e$, when the line $L_3$
is
crossed (at large values of $\e$ not shown in 
Fig.~\ref{fig:planeepsgam}(b)),
 multiplicity disappears.

In contrast to Fig.~\ref{fig:gam_08}(a), the first synchronous solution
$\sigma=0$ in Fig.~\ref{fig:gam_08}(b) looks like arising via 
a second-order
phase transition. 
However a detailed analysis of the situation 
in Fig.~\ref{fig:gam_08}(d) shows that it is not the
case (as was erroneously stated in~\cite{Komarov-Pikovsky-13a}). 
With decrease of parameter $u$ to zero (decrease of $\e$), lines $L_1$ and
$L_2$ come close to each other but they merge only in the point $u=0$ which
corresponds to the  pure second-harmonics Kuramoto model ($\e=0$).
In the Section~\ref{sec:perturb} below, using a
combination of the self-consistent approach and of a perturbative
analysis, we will show that at $L_2$ the dependence of $R_{1,2}$
on coupling strengths $\e$ and $\gamma$ is linear with negative slope,
everywhere except singular points $u=0$ and $u=\pi/2$ which correspond to
the pure cases of second-harmonic and 
first-harmonic Kuramoto models, respectively.

\subsection{Stability properties}

Unfortunately, we cannot perform analytically, and even numerically, a thorough 
stability analysis 
of the 
constructed solutions. The only analytic results we can rely on, are 
outlined in Section~\ref{sec:stis}
stability
calculations
of the asynchronous state $R_{1,2}=0$, yielding instability 
for $\e>\e_{\text{lin}}$ or $\gamma>\gamma_{\text{lin}}$,
and neutral stability due to a continuous spectrum
otherwise~\cite{Strogatz-Mirollo-91,%
Crawford-95,Crawford-Davies-99,Chiba-Nishikawa-11}. 
This conclusion can be easily reproduced numerically, 
see Fig.~\ref{fig:stab}.
However, we could not study in the same manner stability of found self-
consistent
solutions, because these solutions have a singular component 
(delta-function in Eqs.~(\ref{eq:distr_func_1},\ref{eq:distr_func_2})).

Therefore, we checked for stability via direct
numerical simulation of large ensembles (see also \citep{Li-Ma-Li-Yang-14}). 
They follow the theoretically
predicted curves, as show markers in Figs.~\ref{fig:gam_08}(a,b). 
At low values of
$R_{1,2}$
these solutions however can be hardly 
confirmed due to finite-size effects.

In order to study these finite-size effects in the 
vicinity of ``bifurcation points'',
i.e. for small values of the order parameters,
we performed additional simulations with large ensemble size $N=2^{18}=262144$. 
Two theoretical curves with $\sigma=0$ and $\sigma=0.2$ for
$\gamma=0.9\gamma_{lin}$ (Fig.~\ref{fig:gam_08}(b)) have been tested 
for stability.
In each simulation we independently generated random 
distribution of frequencies
for $N=262144$ oscillators and prepared initial conditions according to the 
distribution function,
obtained
from our self-consistent analysis
 at given parameters.
As a result, Fig.~\ref{fig:gam_08}(c) shows the averaged values of $R_{1,2}$
(obtained from the numerical simulation of more than 32 independent runs for
each point).
One can see that the  markers are slightly below the curves,
indicating that on average synchronization is weaker than the 
analytically predicted level. Nevertheless, certain level of
coherence is always present and 
it is in a reasonable agreement with analytically
predicted curves.

\begin{figure}
\centerline{\includegraphics[width=0.49\columnwidth]{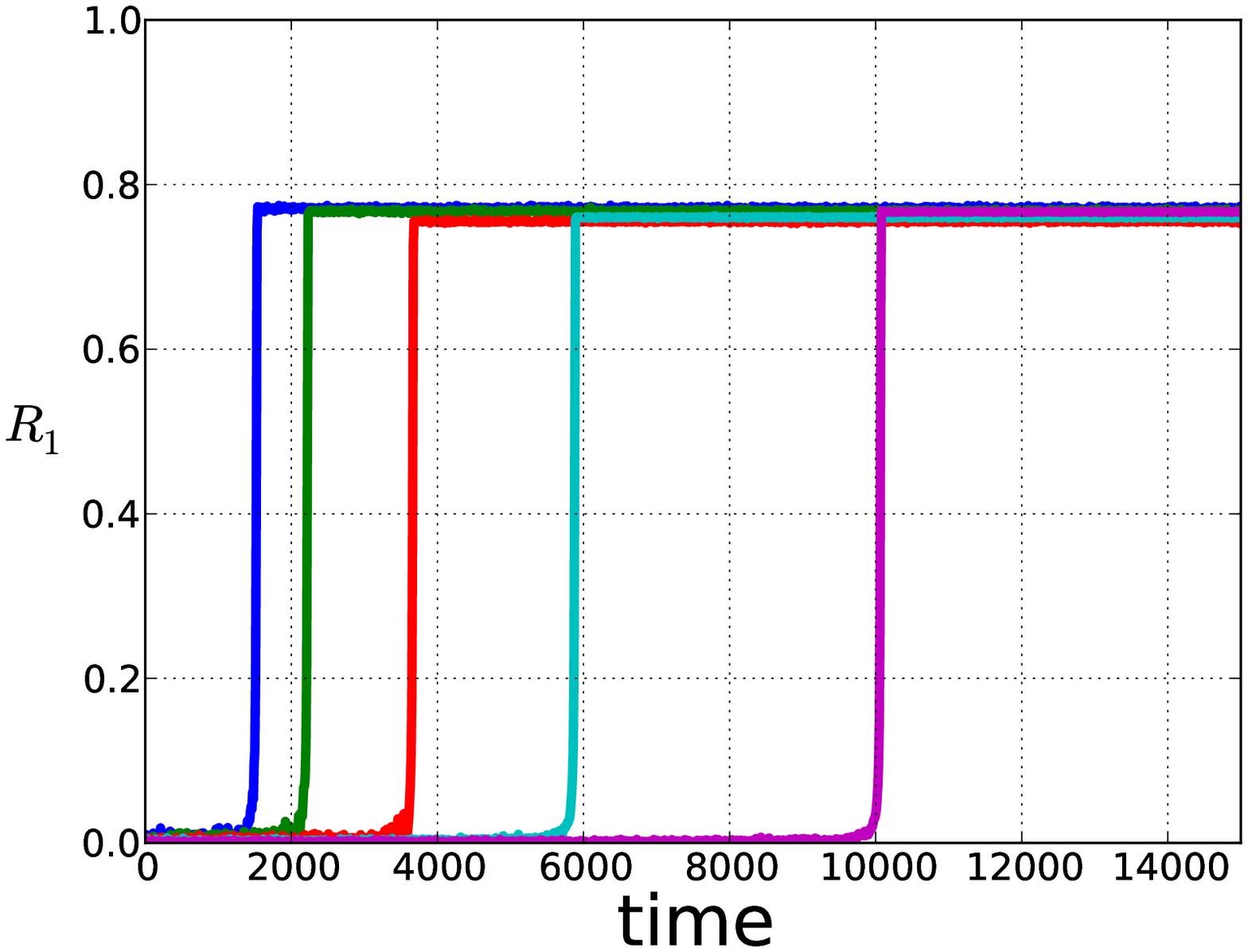}\includegraphics[
width=0.49\columnwidth]{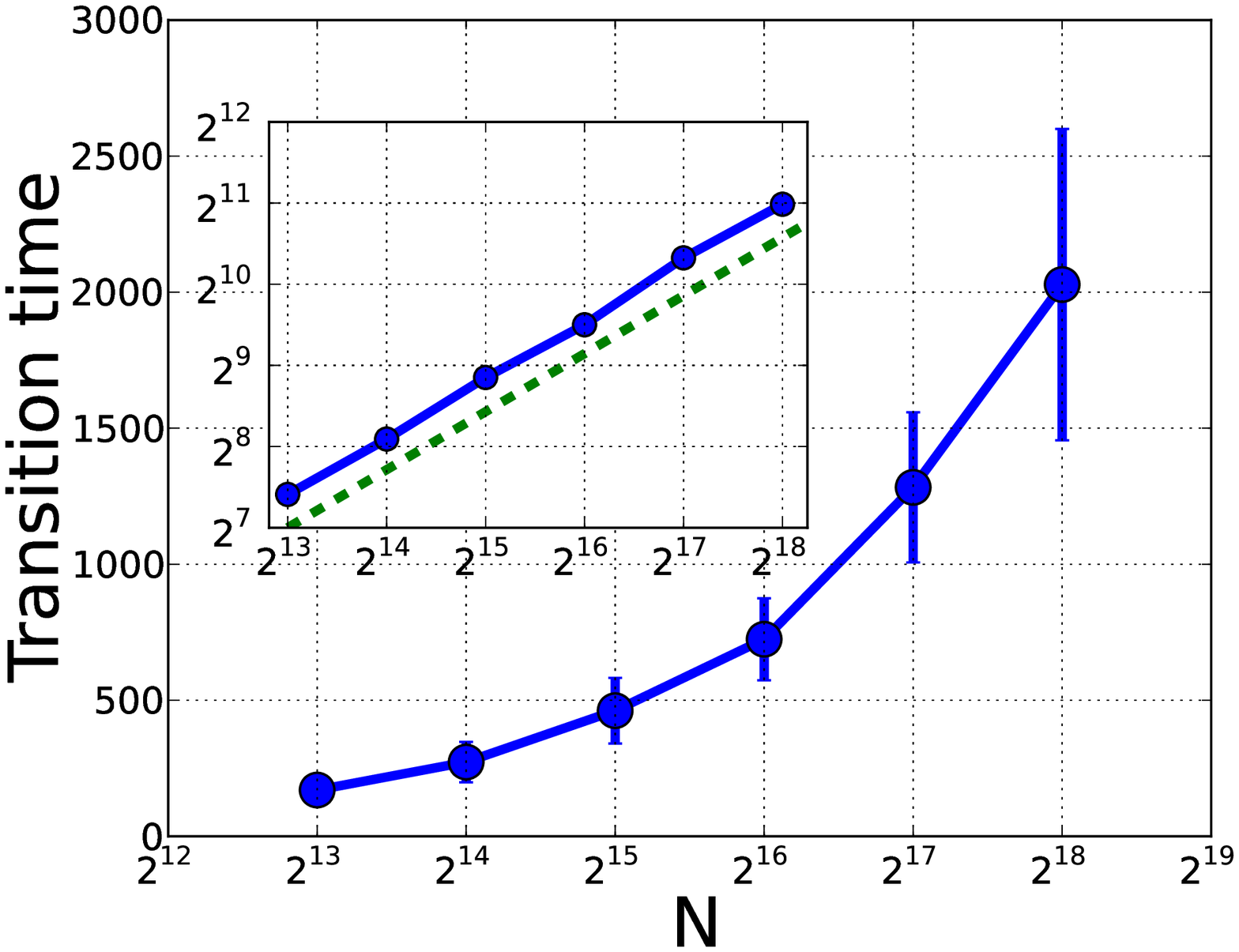} }
\caption{(a) Examples of time evolution of the order parameter $R_1$ in direct
simulations of an ensemble (\ref{eq:km}) for
$\gamma=0.85\gamma_{lin},\e=0.6\e_{lin}$
and different $N$ (from left to right,
$N=5\cdot10^4,10^5,2\cdot10^5,5\cdot10^5,10^6$). 
(b) Averaged transition times from the incoherent state to a synchronous
solution, in dependence on the ensemble size $N$ for $\gamma=0.85
\gamma_{lin}$, $\varepsilon=0.6 \varepsilon_{lin}$. 
Error bars show standard deviations. Each point was obtained from a statistics 
of
128 different simulations. 
Inset shows the same plot in log-log scale. One can see a power low with
exponent $\approx 0.7151$.}
\label{fig:meta}
\end{figure}

Next, we simulated  the linearly neutrally stable asynchronous state, in the
region
beyond the curve $L_2$, where also synchronous solutions exist.
In
simulations this state appears to be only
\textit{metastable}. After a transient, which becomes longer for very large
ensembles, the ensemble
evolves abruptly to one of the synchronous states, we illustrate this
in the Fig.~\ref{fig:meta}(a). Remarkably, the averaged time that the system 
spends in the
vicinity of incoherent metastable state grows as a power low of number of
oscillators $N$ (Fig.~\ref{fig:meta}(b)).

Thus, although the curves in Fig.~\ref{fig:gam_08}(b)
look like for a standard hysteretic 
transition, it is not the case: on line $L_2$ (at point $P$)
the incoherent steady state does not 
become linearly unstable, instead it remains
linearly neutrally stable
in the thermodynamic limit, but is metastable due to finite-size effects. This
neutral 
stability/metastability allows also synchronous states to appear with
{arbitrary}
small amplitudes $R_{1,2}$ 
  (see
 on Fig.~\ref{fig:planeepsgam}(a,b) curve $L_2$ and corresponding curves for
different values
 of $\sigma$, which occupy the whole region on this diagram, and also 
 Fig.~\ref{fig:gam_08}(b)). 
Therefore, the points in Fig.~\ref{fig:gam_08}(b)
where $R_{1,2}$ 
vanish, do not correspond to a usual
bifurcation from an equilibrium, and cannot be described as the points where the
incoherent state becomes linearly unstable. 
While this issue requires further investigation, we attribute it  
to singularity of the appearing states: as one can see from
Eq.~\eqref{eq:distr_func_1},\eqref{eq:distr_func_2}, the density includes 
a combination of delta-functions for any small $R_{1,2}$, similar to the 
Van Kampen modes in plasmas~\cite{VanKampen-55}, while in the
stability analysis~\cite{Crawford-95,Crawford-Davies-99} one operates with
modes which
apparently cannot straightforwardly 
describe constructed singular solutions.

\subsection{Illustration of multi-branch entrainment states}
\label{sec:mult}

\begin{figure}
\centerline{\includegraphics[width=0.45\columnwidth]{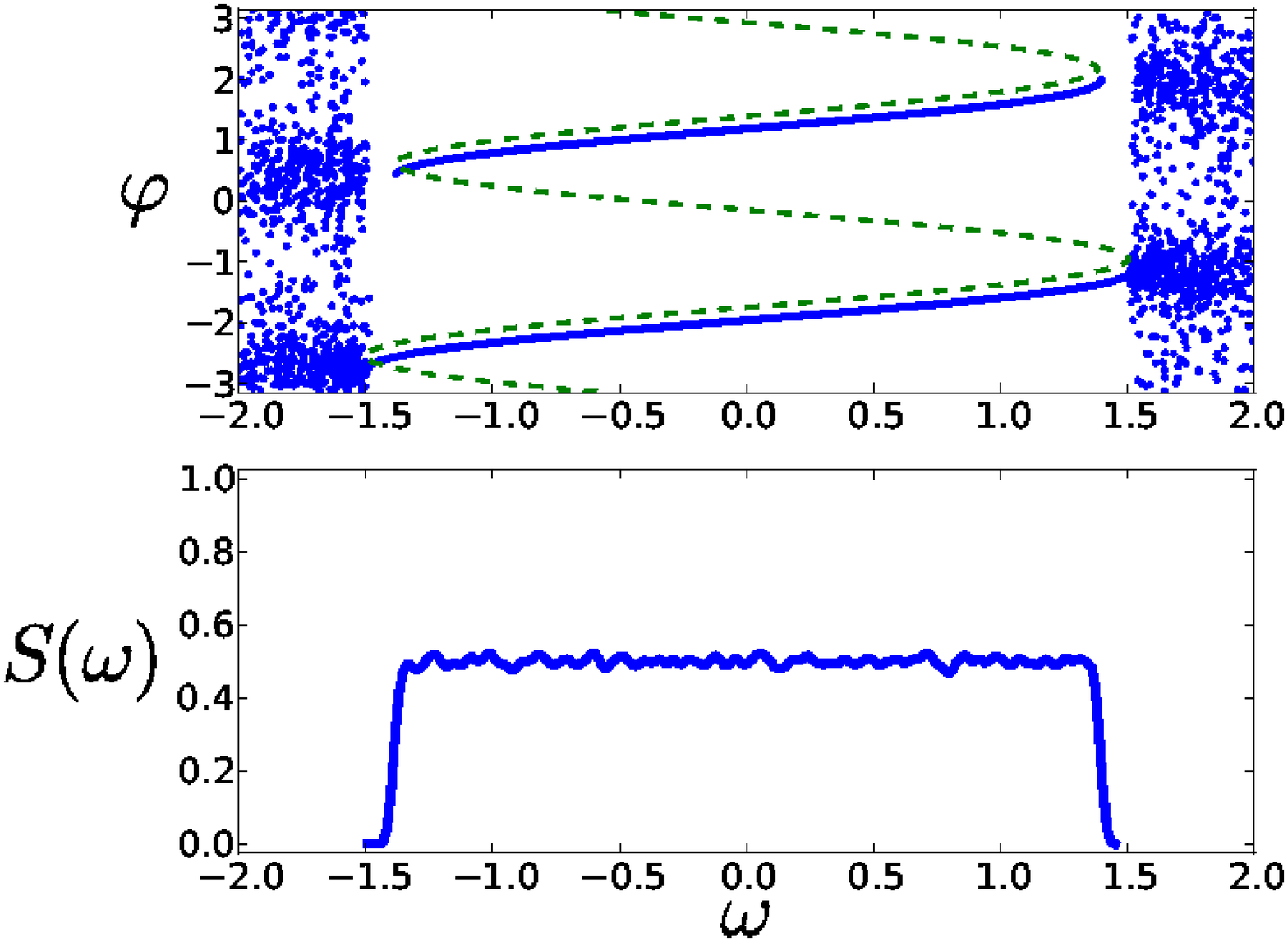}\hfill
\includegraphics[width=0.45\columnwidth]{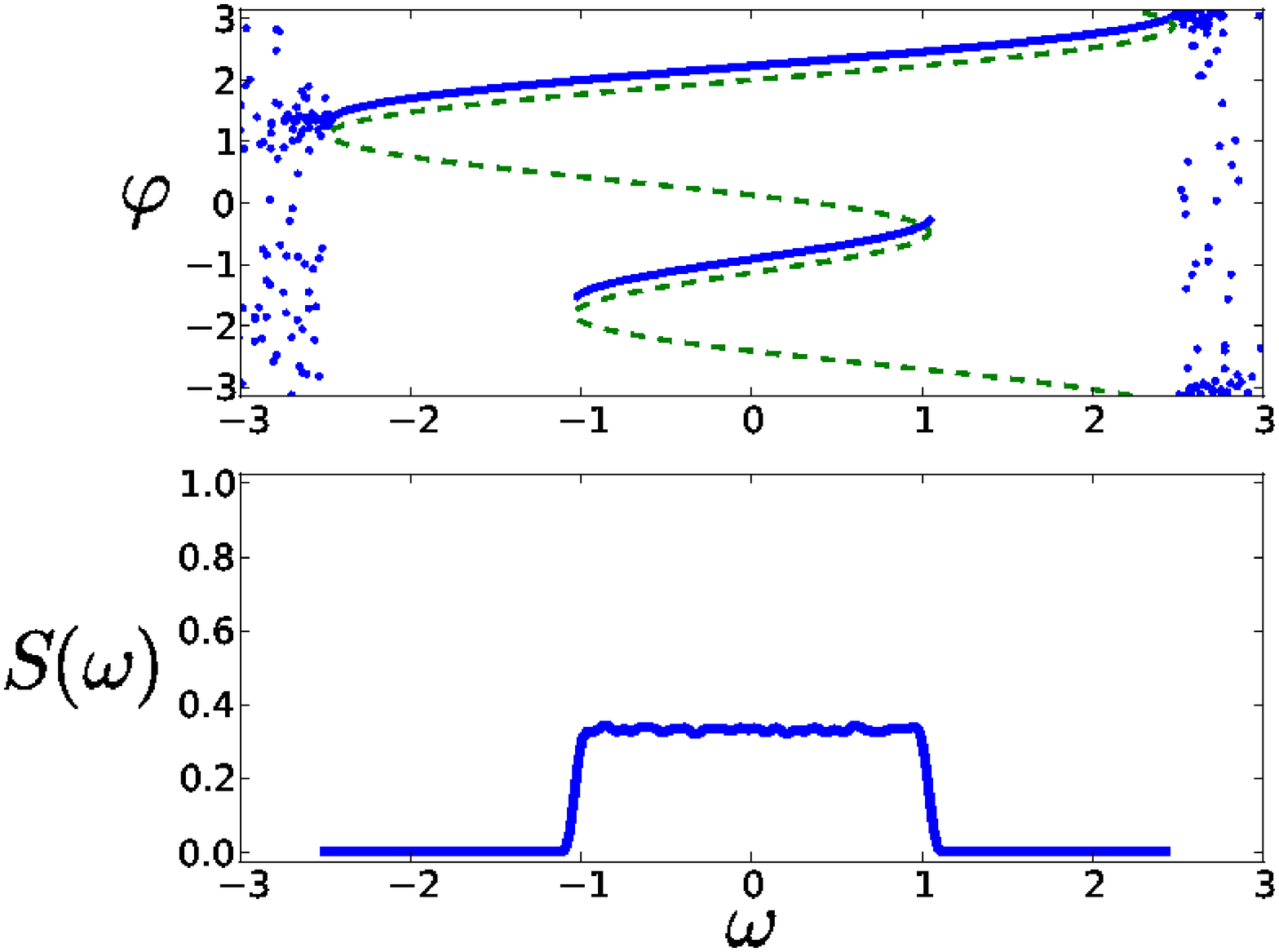} }
\centerline{\includegraphics[width=0.45\columnwidth]{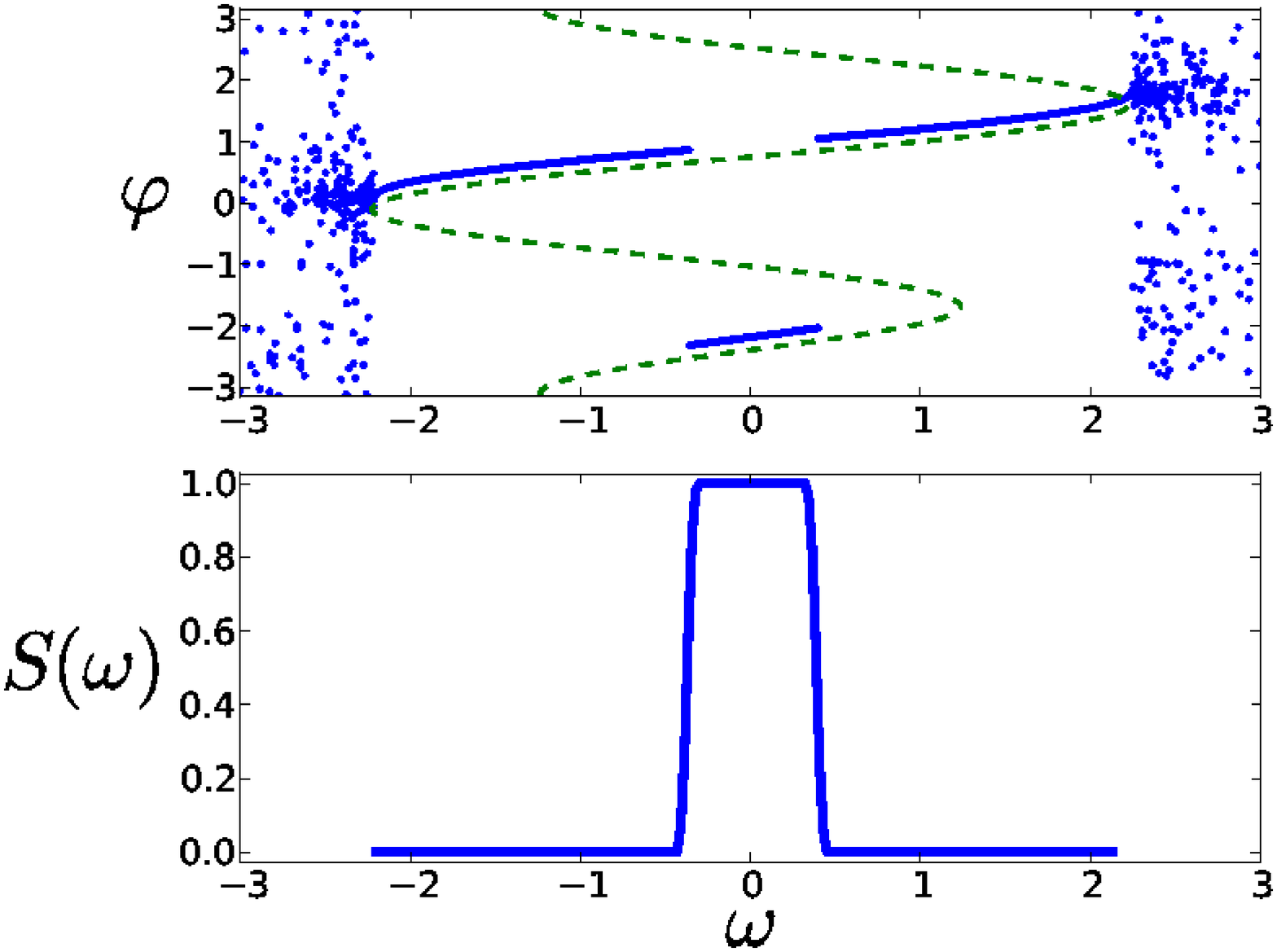}\hfill
\includegraphics[width=0.45\columnwidth]{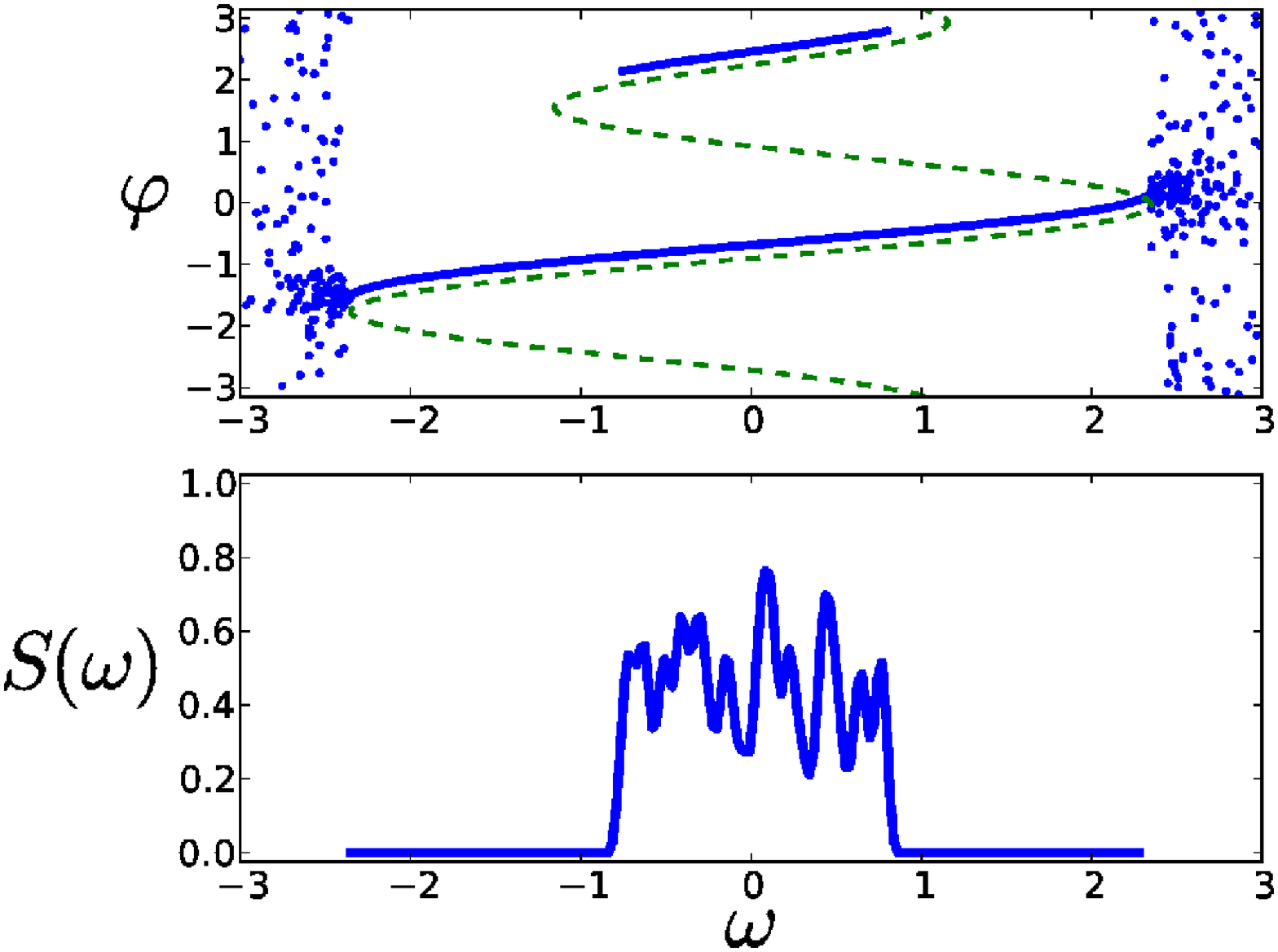} }
\caption{Illustration of multiplicity of  states
($\e=\gamma=1.25\e_{\text{lin}}$,
$N=2\cdot 10^4$).
In all cases one can see two stable branches of locked phases and the
corresponding
coarse-grained indicator function $S(\w)$. }
\label{fig:particles}
\end{figure}

Here we discuss the issue of multiplicity and illustrate different multi-branch
entrainment states~\citep{Daido-95,Daido-96a}. 
As mentioned above, in the thermodynamic limit any 
indicator function $S(x)$ admittable, so that for fixed parameters $\e,\gamma$,
 to a macro-state with given order parameters
$\e,\gamma,R_{1,2}$ belong many
micro-states with different redistributions 
between the stable branches. 
In Fig.~\ref{fig:particles} we show several multi-branch states for a certain 
choice
of coupling parameters. If both branches are occupied, one observes a two-hump
distribution of locked phases which can be also interpreted as   
a two-cluster state (cf. \cite{Kiss-Zhai-Hudson-05}). 

In fact, we can easily estimate the degree of the
multiplicity.
We can view the locked oscillators in the bistability range as ``uncoupled 
spins''. 
Assuming for simplicity that the phases of two branches differ by $\pi$, we
conclude that the order 
parameter $R_2$ does not depend on the ``spin orientation'', i.e. on which
branch they are sitting, while $R_1$ can be interpreted as a 
``magnetization''. Then finding the number of different micro-states  
at prescribed values of the order parameters
reduces to a textbook problem of calculating the entropy $$\mathcal{S}(R_1)=
N_{\text{bist}}\left[-\left(\frac{1-R_1}{2}\right)\ln 
\left(\frac{1-R_1}{2}\right)-
\left(\frac{1+R_1}{2}\right)\ln \left(\frac{1+R_1}{2}\right)\right]$$
for a constant magnetization for $N_{\text{bist}}$ non-interacting spins
(the latter is the number
of locked oscillators 
in the range of bistability; it is less than $N$ but is a 
macroscopic quantity for $R_{1,2}$ not too small). 
Correspondingly, the number of micro-states grows exponentially with the number
of locked oscillators $\sim e^{\mathcal{S}(R_1)}$ (cf.~\cite{Daido-96a}).

\subsection{Competition of the coupling terms}

\begin{figure}[!htb]
\centerline{(a)\includegraphics[width=0.45\columnwidth]{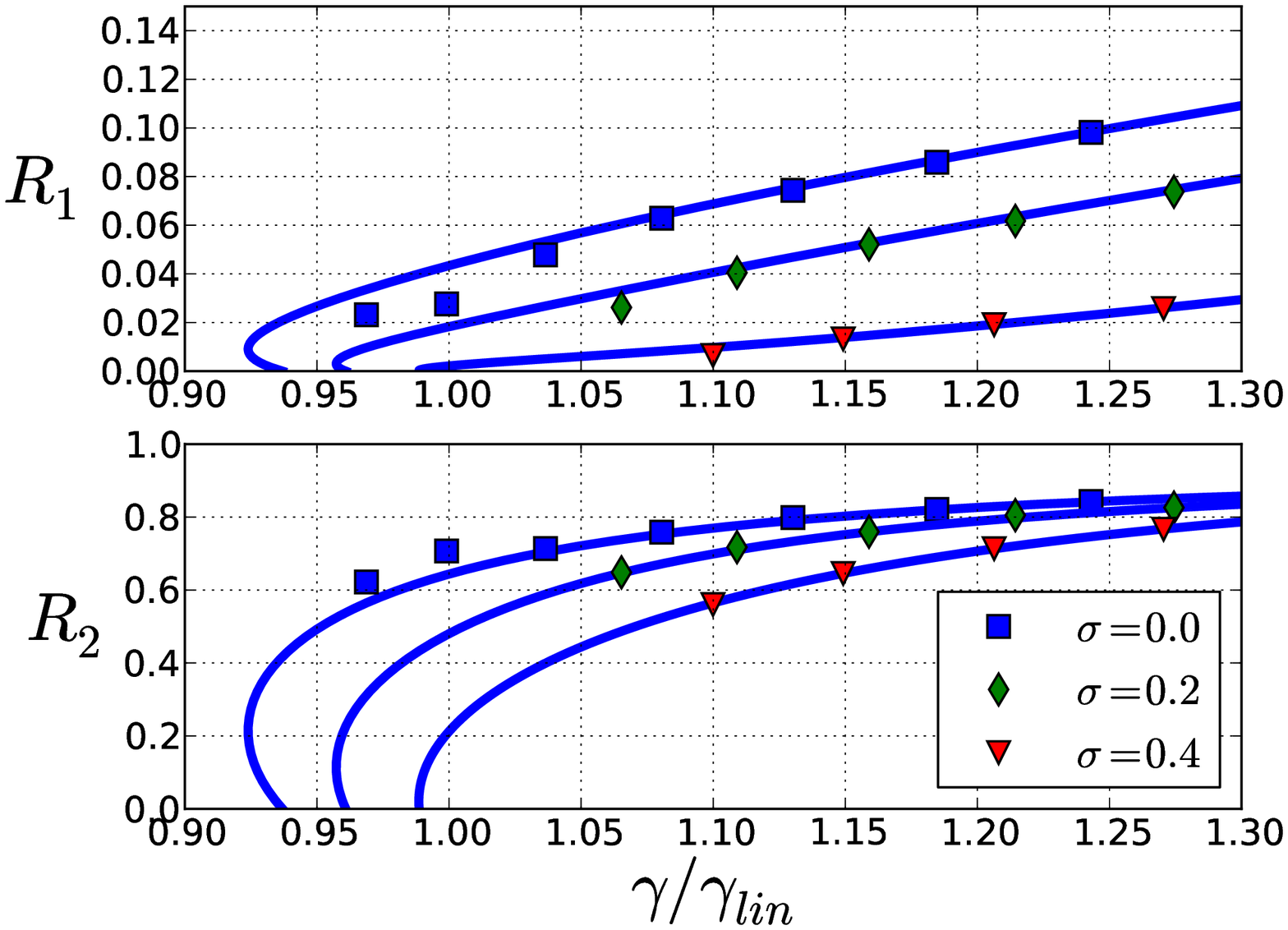} }
\centerline{(b)\includegraphics[width=0.45\columnwidth]{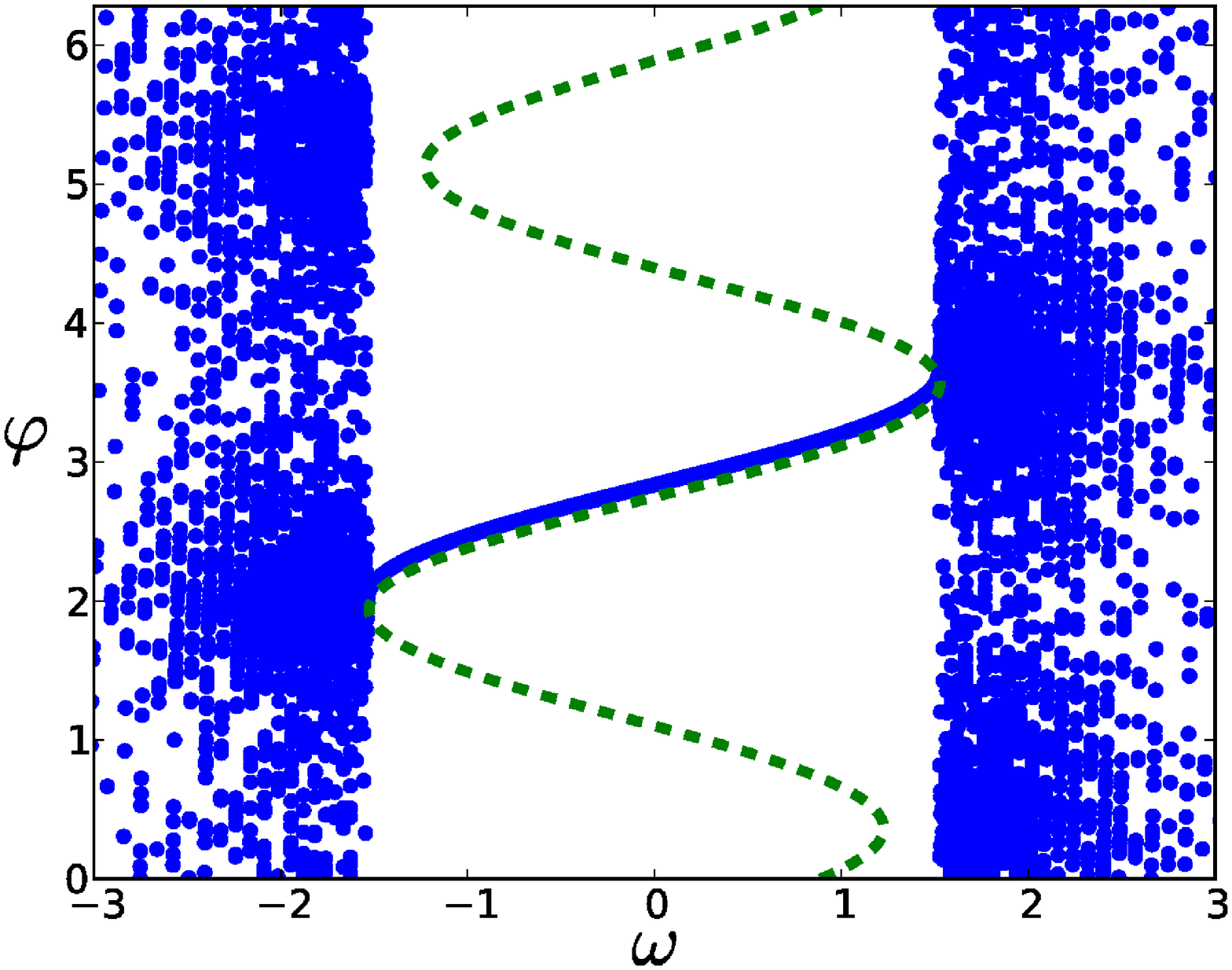}
(c)\includegraphics[width=0.45\columnwidth]{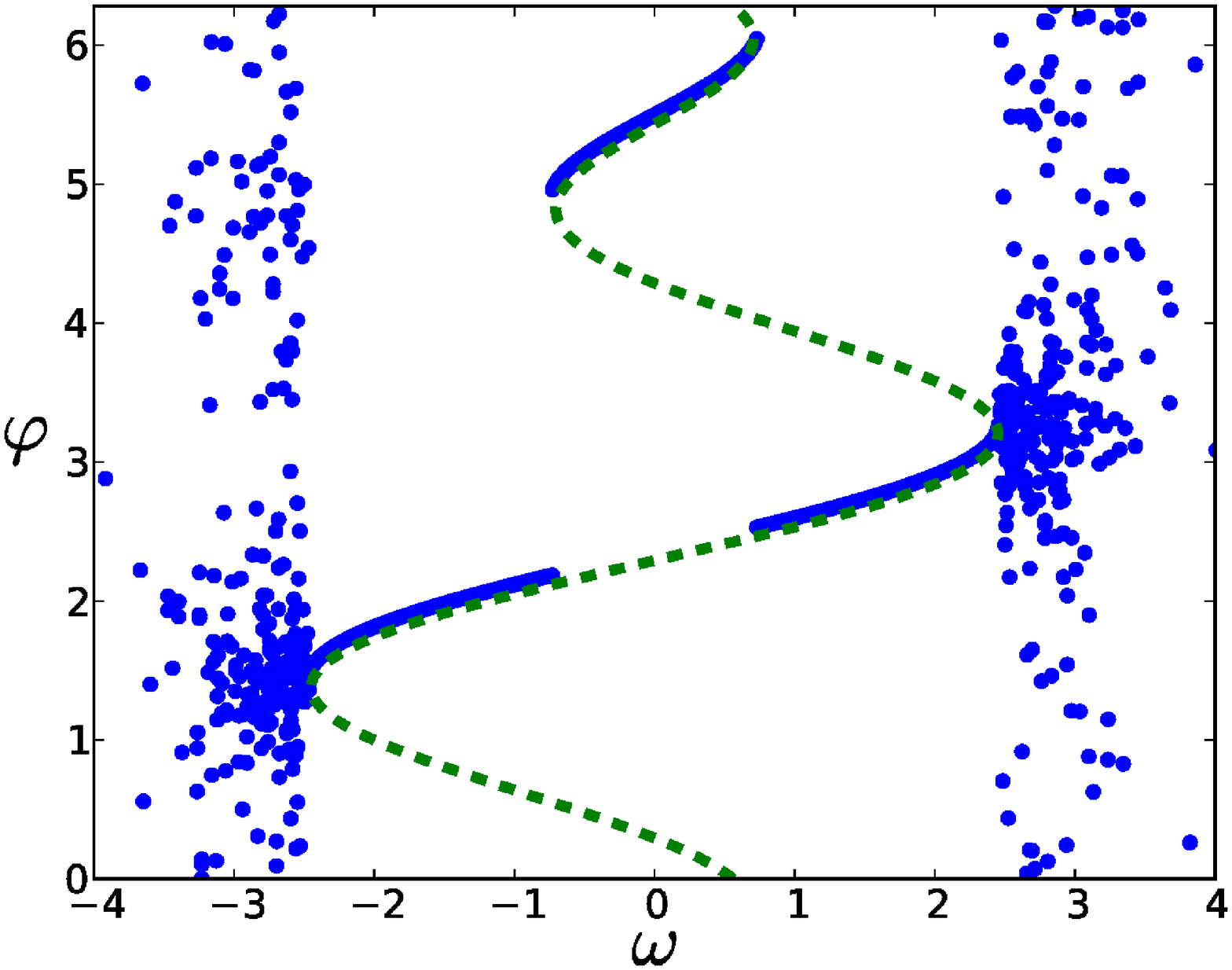} }
\caption{
(a) Dependence of order parameters 
$R_{1,2}$ on coupling strength $\gamma$ at $\e=
-9.29\gamma_{lin}$. 
(b,c) Phases of oscillators versus internal frequencies are plotted. For both
cases $\gamma=1.18\gamma_{lin}$. In the plot (a) $\e=0.16\e_{lin}$, in (b)
$\e=-9.29\e_{lin}$. Markers are result of direct simulation for $N=2\times 10^{4}$.
}
\label{fig_em20}
\end{figure}

A non-trivial consequence of the multi-branch entrainment occurs 
in the region of negative $\e$. 
When $\e<0$,  the
coupling due to the first
mode in the coupling function is repulsive 
(or desynchronizing); it tends to stabilize the incoherent state and to
destroy synchrony. 
With the second harmonic in coupling function, one might expect that 
the repulsion for
large negative $\e$ should be compensated by a 
strong attractive second-harmonic coupling
with large positive $\gamma$, for synchronization in the
system to occur.
However, following the curve $L_1$ in Fig.~\ref{fig:planeepsgam}(a) one can see
that the critical value of $\gamma$ \textit{decreases} and 
tends to some constant value below
$\gamma_{\text{lin}}$ as $\e\to -\infty$. 
This means that the effect of very strong repulsive coupling via the first
harmonic 
can be compensated by a relatively weak synchronizing force $\sim \gamma$.
Figure~\ref{fig_em20}(a) shows dependencies $R_{1,2}(\gamma)$ at $\e =
-9.29\e_{lin}$. 
Remarkably, the presented solutions are characterized by rather low values
of $R_1$. 
The plots of $\phi(\w)$ in Fig.~\ref{fig_em20}(b,c) shed light onto this
effect. 
In the region $\e>0$ the solutions appearing on the line $L_1$ have simple
structure of single-branch entrainment states (Fig.~\ref{fig_em20}(b)). 
On the contrary, in the
region of repulsing first-harmonic coupling $\e<0$,
the appearing solutions represent
two-cluster states with indicator function 
$S=1$, like in Fig.~\ref{fig_em20}(c). 
The oscillators are distributed among the branches 
in such a way that the value
of $R_1$ is minimal, so that effective repulsive force $\e R_1$ 
(see eq.(\ref{eq:bkm_th}))
is sufficiently weak.

\subsection{Non-symmetric solutions}
\label{sec:nonsym}

\begin{figure}
\centerline{\includegraphics[width=0.45\columnwidth]{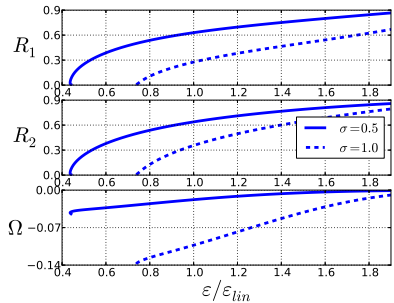} }
\caption{
Dependence of the order parameters $R_{1,2}$ and their frequency
$\Omega$ on coupling strength $\varepsilon$ at $\gamma=0.9\gamma_{lin}$ 
and $\beta_{1,2}=0$ for non-even indicator function: 
$S(x)=\sigma$ for $x<(x^b_1+x^b_2)/2$ (see Fig.~\ref{fig:locked_ph}(b)) 
and $S(x)=0$ otherwise.
}
\label{fig:nonsym}
\end{figure}

Until now we considered the cases where the functions 
$S(x)$ (indicator function) and $g(x)$ (distribution of frequencies) were even 
$S(x)=S(-x)$, $g(x)=g(-x)$.
Such symmetric indicator and frequency distribution functions 
yield solutions with $\beta_{1,2}=0$ and $\Omega=0$ at zero values of parameters 
$z=v=0$ in the self-consistent equations 
(\ref{eq:int1},\ref{eq:self_cons1},\ref{eq:self_cons2}).
However in the general case of non-even $S(x)$ or $g(x)$ zero values of 
$\beta_{1,2}$ correspond to certain non-zero $z$ and $v$.
For example, asymmetric redistribution of oscillators between stable branches 
(a non-even indicator function $S(x)\neq S(-x)$) 
gives rise to a non-zero frequency 
shift $\Omega=Rz\neq 0$ even in the case of $\beta_{1,2}=0$ and symmetric distribution
of frequencies $g$.
The example is presented in the figure~\ref{fig:nonsym} where we use 
$S(x)=\sigma$ for $x<(x^b_1+x^b_2)/2$ (see Fig.~\ref{fig:locked_ph}(b)) 
and $S(x)=0$ otherwise.

\subsection{Perturbative analysis near critical points}
\label{sec:perturb}

In this section we combine the self-consistent approach
(\ref{eq:int2},\ref{eq:self_cons_vz}) with a perturbative analysis, 
to derive the
scaling law of
macroscopic order parameters in the vicinity of bifurcation line $L_2$
(Fig.~\ref{fig:planeepsgam}) where coherent solution appears.
The idea is to consider (\ref{eq:int2},\ref{eq:self_cons_vz}) in the 
limit $R\to
0$ and to find dependence of $R_{1,2}$ on criticalities $(\e-\e_c)$ and
$(\gamma-\gamma_c)$ in this limit of vanishing order parameters.
For simplicity of presentation we will assume below $S(x)=0$ (all oscillators 
are on the
same branch and $\e_c$, $\gamma_c$ are on the curve $L_2$) and shortly discuss
other possibilities at the end of this section. In this case (\ref{eq:int2}) 
reads
\begin{align*}
F_m &= \int_{0}^{2\pi} d\psi \cos m\psi g(Ry)\frac{\partial y}{\partial \psi} +
\int_{0}^{2 \pi}\int_{|x|>x_1}^{\pm\infty} dx d\psi \frac{g(Rx)C'(x)
\cos(m\psi)}{|x-y( u,\psi)|}\\&
\equiv A_m(R,u)+B_m(R,u)\;,\qquad m=1,2\;.
\end{align*}
(Here $C'$ is the normalization constant).
Because 
$g(x)$ is a symmetric one-hump function, its  expansion for small arguments
reads $g(x)=g(0)-G_2
x^2+\ldots$. 
Suppose that $R \ll 1$, 
than the first term in equation for $F_m$ can be represented using
this series for $g$ as
follows:
\begin{equation}
A_m=\int_{0}^{2\pi} d\psi \cos m\psi (g(0)-G_2 R^2 y^2)\frac{\partial
y}{\partial \psi}=A_{m0}-A_{m2}R^2\;.  
\label{eq:A1}
\end{equation}
For calculation of the second term $B_m$ we first compute
$$
\Phi_m(x) = \frac{\int_0^{2\pi} \frac{d\psi \cos(m\psi)}{|x-y(
u,\psi)|}}{\int_0^{2\pi} \frac{d\psi }{|x-y( u,\psi)|}}
$$
With notation $z=1/x$ we get
\begin{gather*}
\Phi_m(z) = \frac{\int_0^{2\pi} \frac{d\psi \cos(m\psi)}{|1-zy(
u,\psi)|}}{\int_0^{2\pi} \frac{d\psi }{|1-zy( u,\psi)|}}\\
=\frac{\int_0^{2\pi}d\psi \cos(m\psi) [1+zy( u,\psi)+z^2 y^2( u,\psi)+\ldots]}
{\int_0^{2\pi}d\psi [1+zy( u,\psi)+z^2 y^2( u,\psi)+\ldots]}\;.
\end{gather*}
Substituting here expression for $y$ we get
\begin{gather*}
\Phi_1(z)=\frac{z^2\pi\sin u\cos u}{2\pi+z^2 \pi}\approx z^2 \frac{1}{2}\sin
u\cos u=z^2 \Phi_{12}\;,\\
\Phi_2(z)=\frac{z^2\pi(-\sin^2 u/2}{2\pi+z^2 \pi}\approx -z^2 \frac{1}{4}\sin^2
u =z^2 \Phi_{22}\;,
\end{gather*}
or in the old notation
\begin{equation}
\Phi_1(x)=x^{-2} \Phi_{12},\quad 
\Phi_2(x)=x^{-2} \Phi_{22},\quad \Phi_{12}=\frac{\sin
u\cos u}{2},\quad \Phi_{22}=-\frac{\sin^2 u}{4}\;.
\label{eq:phi12}
\end{equation}
The last expressions are valid for $x\gg 1$. For small $x$, $\Phi_m$ are bounded
from above $\Phi_m (x) \leq \bar\Phi_m$.

Now we can rewrite the integrals in the expression for $B_m$ as
\begin{gather*}
B_m = \int_{|x|>x_1}^{\pm\infty} dx g(Rx)\Phi_m(x) =2\int_{x_1}^{\infty} dx 
g(Rx)\Phi_m(x)=\\
=2\int_{x_1}^{\infty} dx  g(0)\Phi_m(x)+2\int_{x_1}^{\infty} dx [g(Rx)-
g(0)]\Phi_m(x)\equiv B_{m0}-\tilde B_m\;.
\end{gather*}
To calculate the last term, we divide the integration range into two 
subintervals
\begin{equation*}
\begin{aligned}
\tilde B_m&=2\int_{x_1}^{\infty} dx [g(0)-g(Rx)]\Phi_m(x)\\&=2\int_{x_1}^{R^{-1/6}}
dx [g(0)-g(Rx)]\Phi_m(x)+
2\int_{R^{-1/6}}^{\infty} dx [g(0)-g(Rx)]\Phi_m(x)
\end{aligned}
\end{equation*}
In the first interval we use the upper bound for $\Phi_m$, and because here
$Rx\ll 1$, we use the expansion $g(x)=g(0)-G_2 x^2$:
\begin{equation*}
\begin{aligned}
&2\int_{x_1}^{R^{-1/6}} dx [g(0)-g(Rx)]\Phi_m(x)<2\bar\Phi_m G_2 
R^2\int_{x_1}^{R^{-1/6}} x^2 dx=\\
&2/3\bar\Phi_m G_2 R^2 (R^{-1/2}-x_1^3)
=\mathcal{O}(R^{3/2})
\end{aligned}
\end{equation*}
In the second integral, 
because $x\gg 1$, we use the  expansion (\ref{eq:phi12}) for $\Phi_m(x)$
\begin{gather*}
2\int_{R^{-1/6}}^{\infty} dx
[g(0)-g(Rx)]\Phi_m(x)=2\Phi_{m2}\int_{R^{-1/6}}^{\infty} dx
[g(0)-g(Rx)]x^{-2}=\\
=2\Phi_{m2}R\int_{R^{5/6}}^{\infty} dz [g(0)-g(z)]z^{-2}=\\
2\Phi_{m2}R\int_{0}^{\infty} dz [g(0)-g(z)]z^{-2}-2\Phi_{m2}R\int_{0}^{R^{5/6}}
dz [g(0)-g(z)]z^{-2}\approx \\
\Phi_{m2}R\Gamma -2\Phi_{m2}R G_2 \int_{0}^{R^{5/6}} dz\approx \Phi_{m2}RQ
\end{gather*}
where
$$
Q=2\int_{0}^{\infty} dz [g(0)-g(z)]z^{-2}
$$
characterizes the frequency distribution, 
and we neglected terms having higher orders in $R$.
Summing together we get
$$
B_m=B_{m0}-R Q \Phi_{m2}\;.
$$

Thus, in the leading order, 
we obtain the following expressions for the functions $F_m$:
\begin{equation}
F_m(R,  u)=A_{m0}+B_{m0}-R\Gamma \Phi_{m2}=F_{m0}( u)-RQ \Phi_{m2}( u)\;.
\label{eq:F}
\end{equation}
Here we can immediately  
identify cases where the expansion (\ref{eq:F}) is not sufficient: 
these are situations where $\Phi_{m2}=0$. 
For $u=0$ we have $\Phi_{12}=\Phi_{22}=0$; according to 
Eqs.~(\ref{eq:self_cons_vz}) this corresponds to $\e=0$, i.e. to pure second 
harmonic coupling. For
$u=\pi/2$ only one coefficient vanishes $\Phi_{12}=0$, this corresponds to the 
standard Kuramoto
model with   $\gamma=0$. In both cases the dependencies of the order parameters 
on the coupling
constants follow the square-root law $R_1\sim(\e-\e_{\text{lin}})^{1/2}$, 
$R_2\sim(\gamma-\gamma_{\text{lin}})^{1/2}$~\citep{Kuramoto-84}.

Using general 
expression~(\ref{eq:F}) we can find how the order parameters depend on the 
coupling constants for any
crossing of the critical curve.
Suppose we consider a critical point $\varepsilon_c,\gamma_c$
corresponding to $u_c$, and we choose some direction $q$ of
 crossing
the criticality, so that $ u= u_c+q R$. 
Then
\begin{gather*}
\varepsilon=\frac{\sin u}{F_{10}( u)-R\Gamma \Phi_{12}( u)}=
\frac{\sin u_c +\cos u_c q R}{F_{10}( u_c)+(F'_{10} q-\Gamma \Phi_{12}(
u_c))R}=\\
\frac{\sin u_c}{F_{10}( u_c)}+R[q\frac{\cos u_c}{F_{10}( u_c)}-\frac{\sin u_c
(F'_{10} q-\Gamma \Phi_{12}( u_c))}
{F_{10}^2( u_c)}]=\\
=\varepsilon_c+\varepsilon_1(q) R\\ \\
\gamma =\frac{\cos u}{F_{20}( u)-R\Gamma \Phi_{22}( u)}=
\frac{\cos u_c -\sin u_c q R}{F_{20}( u_c)+(F'_{20} q-\Gamma \Phi_{22}(
u_c))R}=\\
\frac{\cos u_c}{F_{20}( u_c)}+R[q\frac{-\sin u_c}{F_{20}( u_c)}-\frac{\cos u_c
(F'_{20} q-\Gamma \Phi_{22}( u_c))}
{F_{20}^2( u_c)}]=\\
=\gamma_c+\gamma_1(q) R
\end{gather*}
This yields
\begin{equation}
R_m=\frac{F_{m0}( u_c)}{\varepsilon_1(q)}(\varepsilon-\varepsilon_c)=
\frac{F_{m0}( u_c)}{\gamma_1(q)}(\gamma-\gamma_c)
\label{eq:lin_sc_gen}
\end{equation}

Choosing parameter $q=q_0$ in such a way that $\gamma_1(q_0)=0$ we have:
\begin{equation}
R_m=\kappa_m^\e(\e-\e_c),\ \gamma\equiv \gamma_c;
\label{eq:lin_sc_eps}
\end{equation}
$\gamma_1(q_0)=0$ implies that:
$$
q_0 = \frac{\cos u_c\Gamma\Phi_{22}( u_c)}{\sin u_c F_{20}( u_c)+\cos u_c
\frac{\partial F_{20}}{\partial q}}
$$

The same for $\e_1(q_1)=0$:
\begin{equation}
R_m=\kappa_m^\gamma(\gamma-\gamma_c),\ \e\equiv\e_c
\label{eq:lin_sc_gam}
\end{equation}
with
$$
q_1 = \frac{\sin u_c\Gamma\Phi_{12}( u_c)}{\sin u_c\frac{\partial F_{10}(
u_c)}{\partial q}-F_{10}( u_c)\cos u_c}
$$
Here we denote
\begin{equation}
\kappa_m^\e(u_c) = \frac{F_{m0}( u_c)}{\e_1(q_0)},\ \kappa_m^\gamma(u_c) =
\frac{F_{m0}( u_c)}{\gamma_1(q_1)}
\label{eq:kappa}
\end{equation}

Equations~(\ref{eq:lin_sc_gen},\ref{eq:lin_sc_eps},\ref{eq:lin_sc_gam}) show
that generally the order parameters $R_{1,2}$  scale \textit{linearly} at the
``bifurcation points'', in contradistinction to the situations $\e=0$ 
and $\gamma=0$, see
also~\citep{Daido-94} for the first discovery of this scaling.

\begin{figure}
\centerline{\includegraphics[width=0.49\columnwidth]{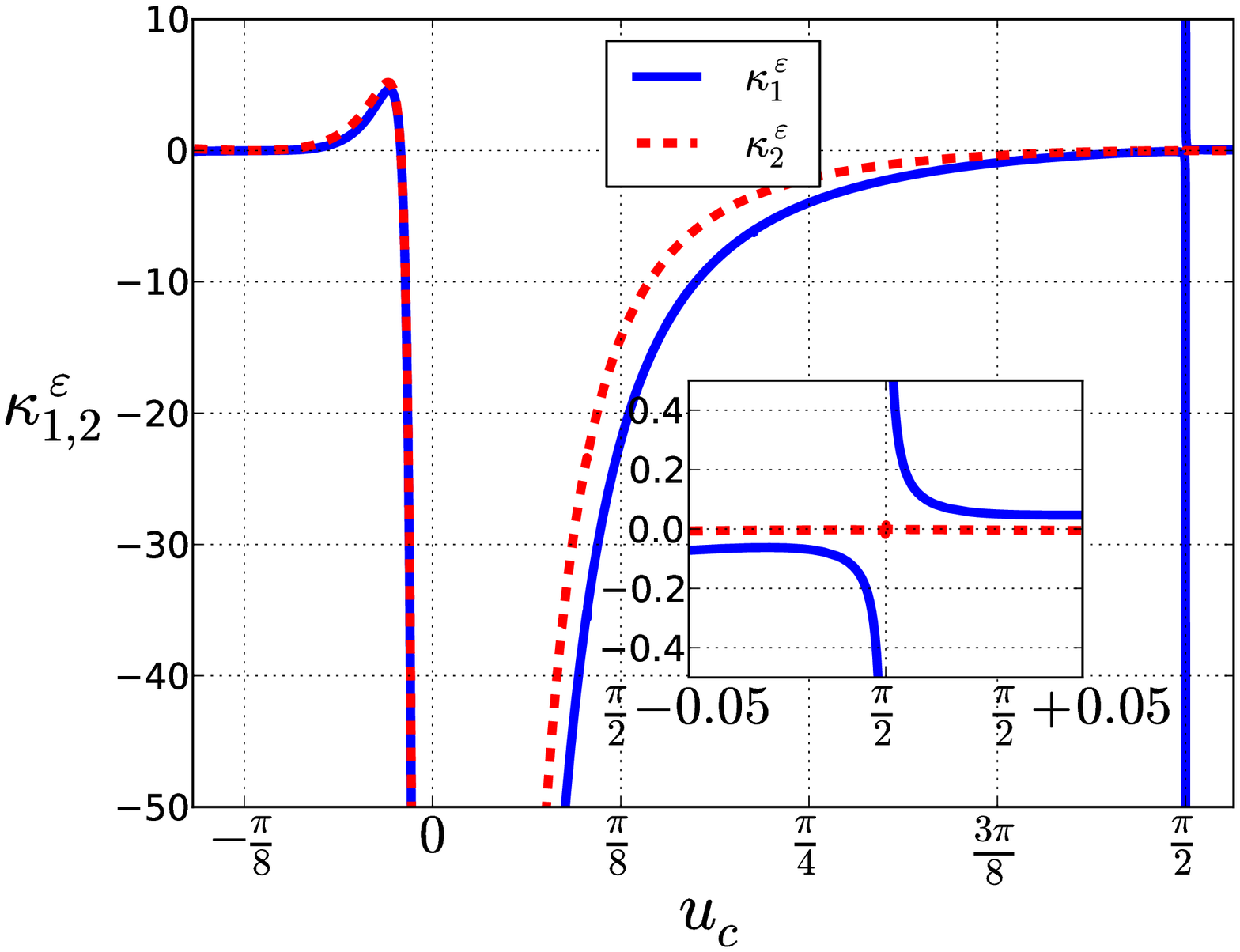}
\includegraphics[width=0.49\columnwidth]{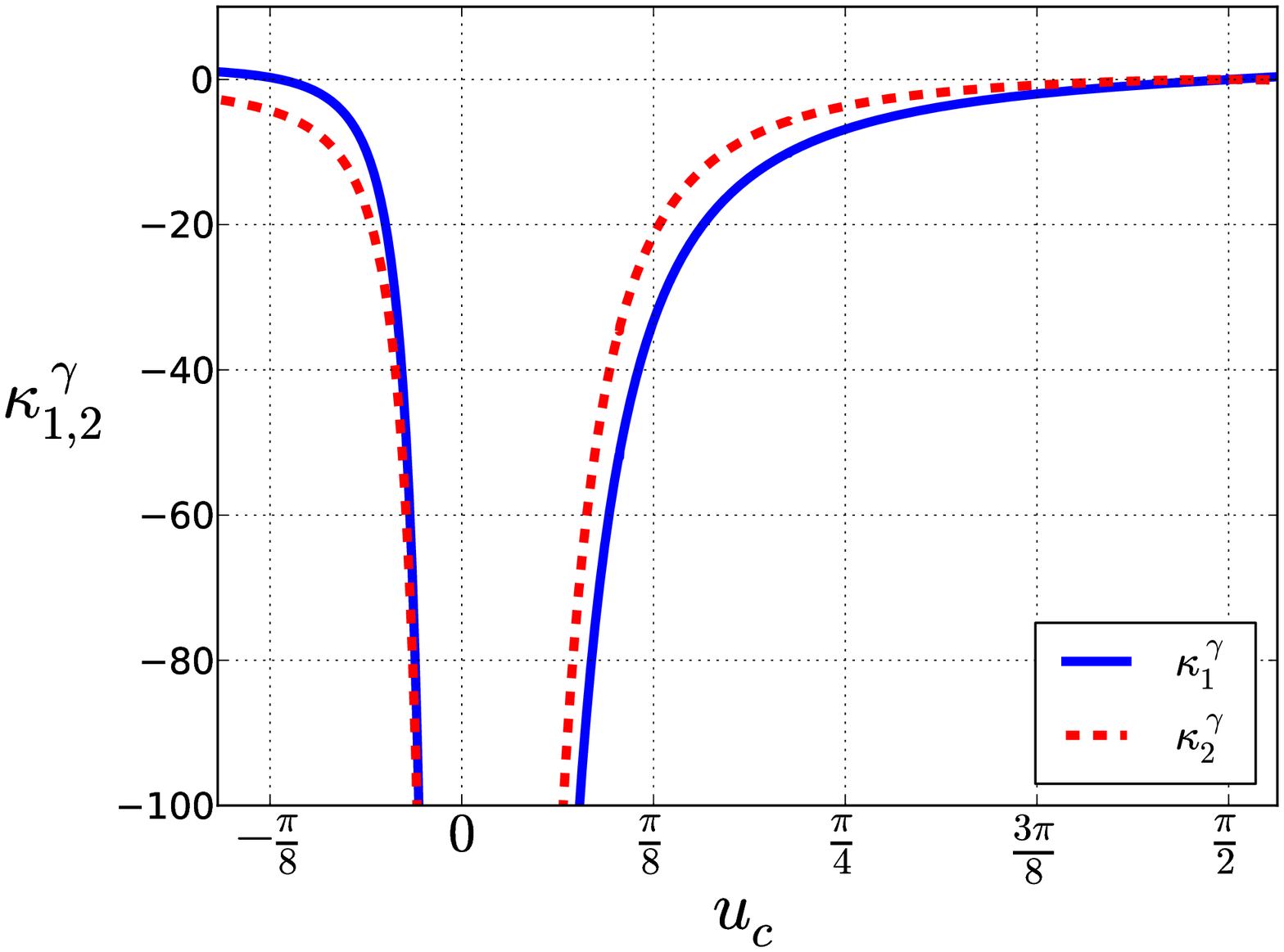} }
\caption{Dependences of $\kappa^{\e,\gamma}_{1,2}$ on $ u_c$.}
\label{fig:kappa}
\end{figure}

For the Gaussian distribution of frequencies $g(\omega) =
\frac{1}{\sqrt{2\pi}}e^{-\frac{x^2}{2}}$ the constant $Q$ can be evaluated
explicitly and it is equal to one.
In the latter case calculations of~(\ref{eq:kappa}) show (Fig.~\ref{fig:kappa})
that $\kappa^{\e,\gamma}_{1,2}(u_c)$ are finite and non-zero everywhere except
for
mentioned above singular points $u_c=0$ and $u_c=\pi/2$, 
which correspond to the one-harmonic
Kuramoto model where the transition has a continuous second-order type form.

\section{Asymmetric coupling function}

\label{sec:asym}

\begin{figure}[!htb]
\centerline{\includegraphics[width=0.5\columnwidth]{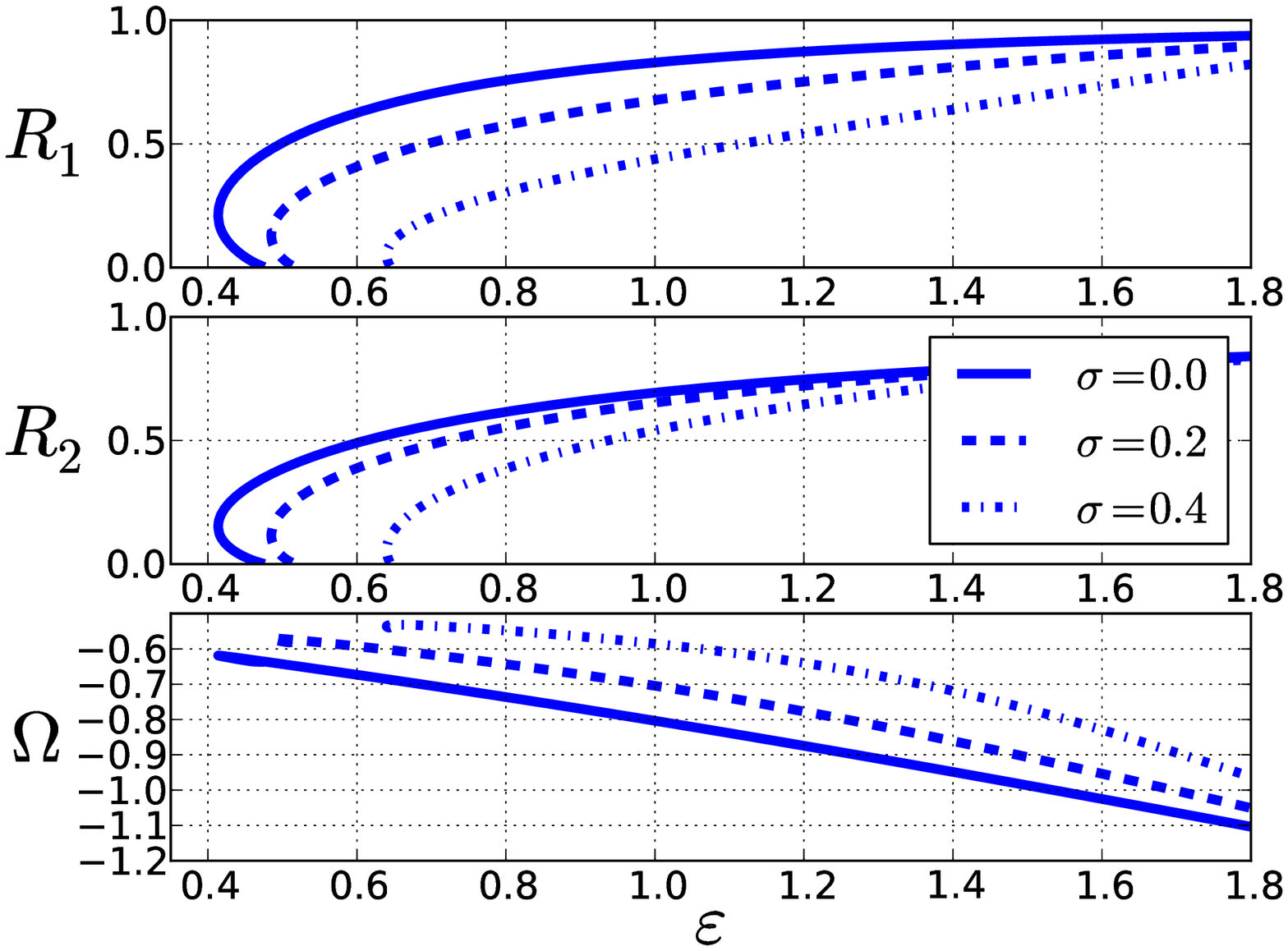}
\includegraphics[width=0.5\columnwidth]{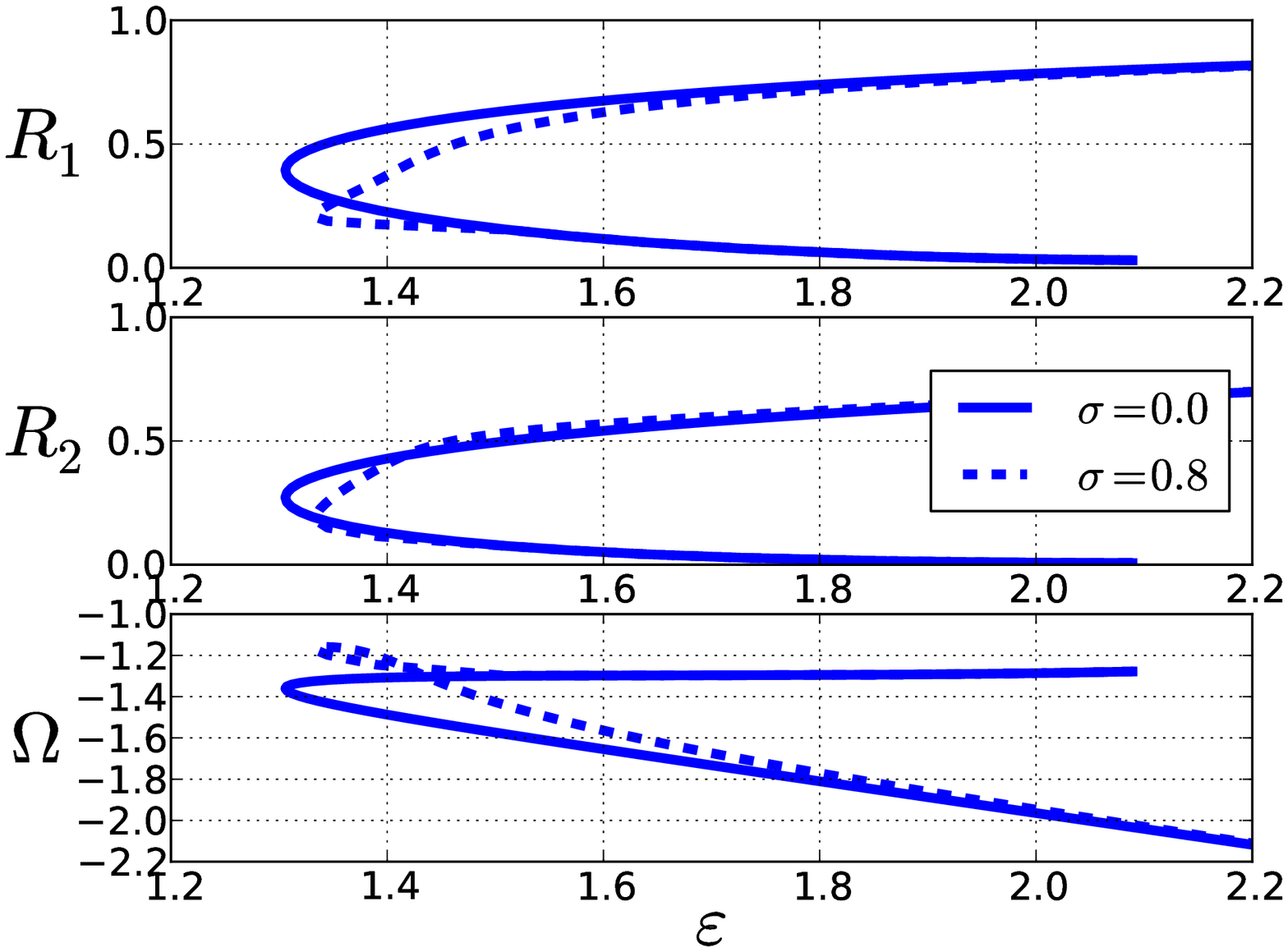} }
\caption{Dependences of the order parameters $R_{1,2}$ and their 
frequency $\Omega$
on coupling strength $\e$ at fixed values of $\beta_{1,2}$ and $\gamma = 1.5$.
In the panel (a) $\beta_{1,2} = \pi/8$, in the panel 
(b) $\beta_{1,2}=\pi/4$. Here no normalization on the linear 
stability thresholds is performed.}
\label{fig_p4}
\end{figure}

In this section we present several examples of application 
of our general theory for
calculation of uniformly rotating synchronous states for the case of
non zero phase shifts $\beta_{1,2}$ in the coupling function, see 
Eqs.~(\ref{eq:self_cons1},\ref{eq:self_cons2}). Here
the number of control parameters ($\e,\gamma,\beta_1,\beta_2$) is large, thus
we do not perform a comprehensive analysis but just illustrate
applicability of the method.

The main general feature at non-zero phase shifts $\beta_1,\beta_2$ is 
a general appearance of a frequency shift
$\Omega$, so that coherent solutions rotate with the frequency different from 
the
mean frequency of the distribution $g(\omega)$. 
Figure \ref{fig_p4} shows dependences of the order parameters
$R_{1,2}$ and of frequency $\Omega$ on coupling constants $\e$
and $\gamma$, for fixed values of $\beta_{1,2}=\pi/8$ (Fig.~\ref{fig_p4}(a)) and
$\beta_{1,2}=\pi/4$ (Fig.~\ref{fig_p4}(b)). These curves have been obtained 
from Eqs.~(\ref{eq:self_cons1},\ref{eq:self_cons2}) 
by adjusting free parameters $\mathbf{P}$
to achieve the given values of $\beta_1,\beta_2$/

Another interesting example is motivated by work by
Hansel~\textit{et.al.}~\citep{Hansel-93} 
In this paper the authors consider an ensemble of \textbf{identical} 
(with equal natural
frequencies) phase oscillators with a bi-harmonic coupling function. At
$\pi/3<\beta_1<\pi/2$, $\beta_2 = \pi$, $\varepsilon/\gamma = 4$ the authors
describe slow periodic oscillations of the order parameters and show that
these variations arise due to a 
closed heteroclinic cycle in the phase space
of the model. In order to model identical oscillators in our setup, one has to
consider a delta-distribution of 
frequencies $g(\omega)=\delta(\omega)$. However,
we have normalized the width of this distribution to one. Because normalization 
of frequencies
is equivalent to normalization of time, in our approach the limit of 
identical oscillators corresponds
to the 
limit $\e,\gamma\to\infty$ at a fixed width of the distribution $g(\omega)$.
Thus, we applied our method for the 
parameters $\beta_{1,2}$ as in~\citep{Hansel-93},
for very large values of the coupling constants.  

\begin{figure}[!htb]
\centerline{(a)\includegraphics[width=0.5\columnwidth]{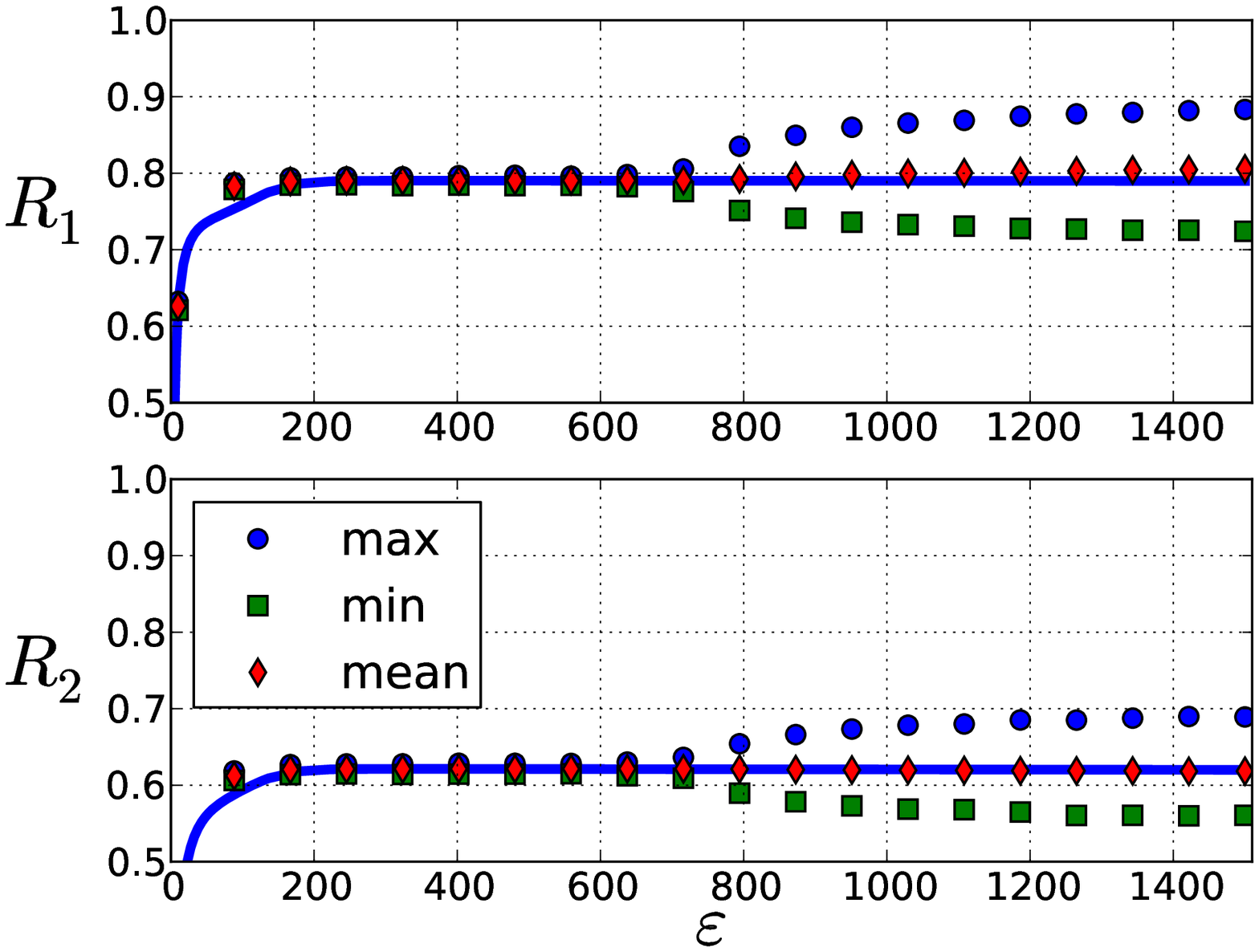} }
\centerline{(b)\includegraphics[width=0.42\columnwidth]{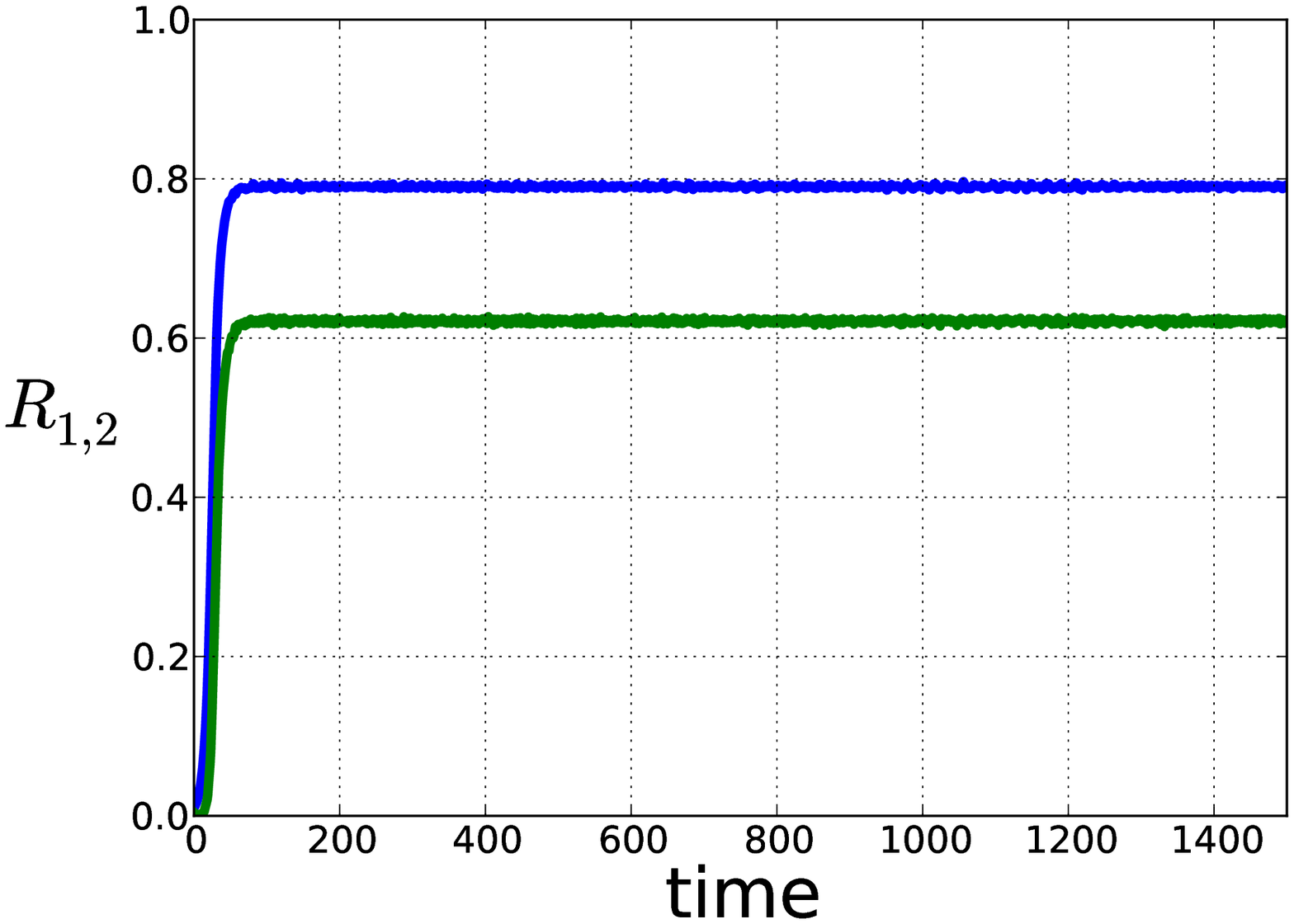}\hfill
(c)\includegraphics[width=0.42\columnwidth]{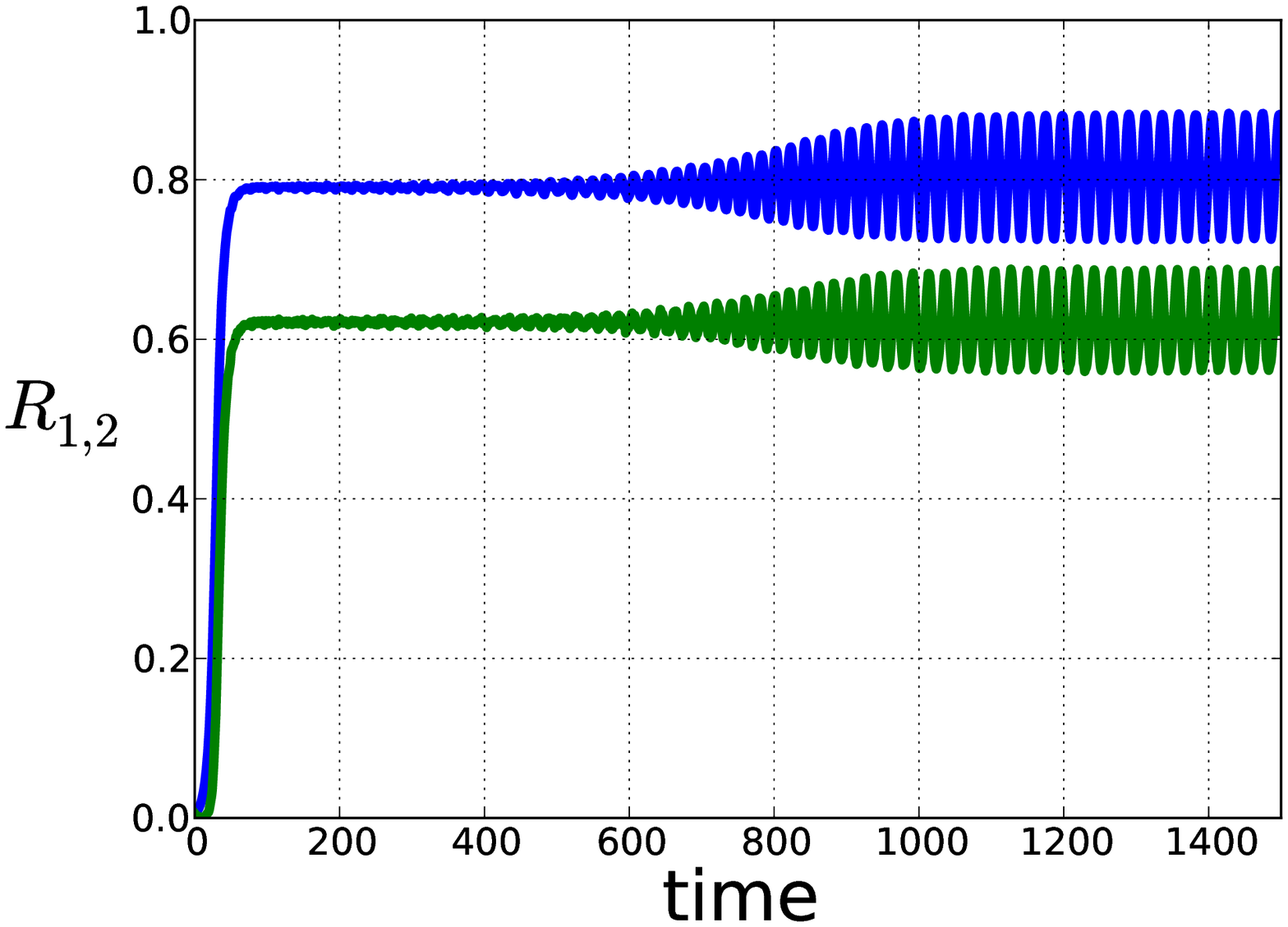} }
\caption{(a) Solutions of self-consistent equations
(\ref{eq:int1}-\ref{eq:self_cons2}) at $\beta_1 = \pi/2.5$, $\beta_2=\pi$ and
$\varepsilon/\gamma = 4$ are shown. 
Markers (showing maximum, minimum and mean values of
$R_{1,2}$ calculated from time series after some transient period)
depict results of direct numerical simulation of equations (\ref{eq:km}) at the
same parameters for $N=2\times 10^{4}$. 
The stationary state loses stability at a large coupling strength $\varepsilon
\approx 650$, beyond which stable oscillations appear. (b,c) 
Time series of $R_{1,2}$ at
different coupling strength are presented: in the panel (b) $\e=323$, in (c) $\e
= 1420$.}
\label{fig_heter}
\end{figure}

Figure~\ref{fig_heter}(a) shows the solutions of
equations~(\ref{eq:int1}-\ref{eq:self_cons2}) at $\beta_1 = \pi/2.5$ $\beta_2 =
\pi$ and $\varepsilon/\gamma=4$, together with the results
of direct numerical simulations
of a large ensemble with $N=2\times10^4$.
At small values of coupling ($\varepsilon<650$), the stationary state obtained
from our self-consistent approach is reproduced by direct numerical
simulations of (\ref{eq:km}) (the time series is shown in
Fig.~\ref{fig_heter}(b)).
At larger values of the coupling, this 
stationary solution loses stability via (presumably) a
supercritical Andronov-Hopf
bifurcation at which slow oscillatory variations of the order parameters
appear (Fig.~\ref{fig_heter}(c)). This example shows that while we always can 
find 
a uniformly rotating solution with constant order parameters, this solution
can be unstable in some parameter range, where a more complex dynamics 
establishes.

\section{Conclusion}
\label{sec:concl}

In this paper we have described nontrivial synchronous states that
appear in
the Kuramoto model
with a bi-harmonic coupling function. 
Here we summarize essential novel features compared to the standard Kuramoto 
setup.
\begin{enumerate}
\item Due to a possibility to have two stable
branches of phase-locked
oscillators, one observes a multi-branch locking with a multiplicity of micro-
states~\citep{Winfree-80,Daido-96a}.   On the macro-level,
this multiplicity 
manifests itself as existence of a whole range of possible order parameters for
given coupling 
constants. We have incorporated this multiplicity of multi-branch states
into an analysis of self-consistent equations for the order parameters,
and presented a general analytic solution. 
\item Appearance of the synchronous states is not related to 
a standard bifurcation, as the
asynchronous state does not change its neutral linear stability. We have found 
domains on the
plane of basic coupling constants for the existence of such solutions, for 
different
 distributions of the locked phases between the 
 branches (Fig.~\ref{fig:planeepsgam}).
\item When a synchronous state is present, numerical experiments with finite 
ensembles
show that the asynchronous state lives a finite 
time that 
scales like
 $T\sim N^{0.7}$, after which an abrupt transition to 
 synchrony occurs. Similarly, we checked numerically stability 
 of the states with 
 single- and multi-branch entrainment through simulations of finite ensembles 
 (Fig.~\ref{fig:gam_08}).
\item At asymmetric distribution between the branches, the frequency of the
order parameters deviates from the central frequency of the distribution, even
if the latter and the coupling are symmetric.
\end{enumerate}

Below we outline some open questions deserving further analysis.
In the case of a general multi-harmonic coupling function $\Gamma$, one can
expect existence of more than two stable brunches for
oscillators at a particular frequency, with
 more possibilities for different redistributions of the oscillators phases.
Another feature not addressed in this paper is related to a possibility 
of non-standard transitions to synchrony for particular distributions of the 
natural frequencies, similar to the analysis presented in 
Ref.~\cite{Omelchenko-Wolfrum-12} for the one-harmonic coupling.
Detailed theoretical 
understanding of stability of the asynchronous states constructed
via the self-consistency approach in this paper, is still missing. 
Finally, noise regularizes the multiplicity of
the micro-states and turns  neutral stability into an asymptotic 
one~\cite{Aonishi-02,Komarov-Pikovsky-13a}; these effects will be discussed in 
details 
elsewhere~\cite{Komarov-Pikovsky-unpublished}.

\section*{Acknowledgements}
M. K. thanks Alexander von Humboldt Foundation for support. 
We acknowledge useful discussions with G. Bordyugov, R. Toenjes, and useful
comments of S. Strogatz.

\section*{Appendix}
\label{sec:appA}

Let us consider a system of $N$ pendulums (with mass $m$ and 
length $l$, described by angles $\theta_j$) 
suspended on a beam of mass $M$, which
 can move
vertically (axis $y$) and horizontally (axis $x$) without rotation. 
These motions 
are controlled by two 
springs $k_x$ and $k_y$. This conservative system is described by the 
Lagrangean (cf. 
\cite{Kapitaniak_etal-12,Czolczynski_etal-13})
\begin{gather*}
L = \frac{M}{2}\left( \dot{x}^2 + \dot{y}^2\right) +
\frac{m}{2}\sum_j \left(\dot{x}^2+\dot{y}^2+l\dot{x}\dot{\theta_j}\cos\theta_j -
l\dot{y}\dot{\theta_j}\sin\theta_j+l^2\dot{\theta_j}^2  \right) +\\ 
+mgl\sum_j \cos\theta_j+gy\left(Nm+M\right)-\frac{k_xx^2}{2}-\frac{k_yy^2}{2}
\end{gather*}
The equations are two equations for the degrees of freedom of the 
beam (where we shift $y$ to the steady position
$g(Nm+M)/ky$), and for each pendulum:
\begin{gather*}
(M+Nm)\ddot{x}+k_xx =\sum_j
-\frac{ml}{2}\ddot{\theta_j}\cos\theta_j+\sum_j\frac{ml}{2}\dot{\theta_j}^2\sin\theta_j\\
(M+Nm)\ddot{y}+k_yy =\sum_j
\frac{ml}{2}\ddot{\theta_j}\sin\theta_j+\sum_j\frac{ml}{2}\dot{\theta_j}^2\cos\theta_j\\
ml^2\ddot{\theta_j}+mgl\sin\theta_j = \frac{ml}{2}\ddot{y}\sin\theta_j -
\frac{ml}{2}\ddot{x}\cos\theta_j
\end{gather*}
In order to model self-sustained oscillations of the pendulum clocks,
we add dissipation terms ($\sim \gamma_{x,y}$) to beam equations, and
van-der-Pol-type self-exciation terms $\sim\sigma$, together 
with cubic saturation,
 to the pendula dynamics.  
In the case of small deviations $\theta_{1,2}$  (i.e. for $\sigma/rho\ll 1$) 
we have:
\begin{gather}
(M+Nm)\ddot{x}+\gamma_x\dot{x}+k_xx =
\sum_j-\frac{ml}{2}\ddot{\theta_j}+
\sum_j\frac{ml}{2}\dot{\theta_j}^2\theta_j\label{eq:aps1}\\
(M+Nm)\ddot{y}+\gamma_y\dot{y}+k_yy =
\sum_j\frac{ml}{2}\ddot{\theta_j}\theta_j+
\sum_j\frac{ml}{2}\dot{\theta_j}^2\label{eq:aps2}\\
\ddot{\theta_j}-(\sigma-\rho \theta_j^2)\dot{\theta_j}+\w^2\theta_j = 
\frac{1}{2l}\ddot{y}\theta_j -
\frac{1}{2l}\ddot{x}\label{eq:aps3}
\end{gather}
where $\w^2=g/l$.

For small $\sigma\ll \omega$ we can apply the averaging (van der Pol) 
method. We will 
seek for a
solution of the form:
$$
\theta_j = A_je^{i\w t}+A^*_je^{-i\w t},\quad 
\dot{\theta}_j = i\w (A_je^{i\w t}-A^*_je^{-i\w t})
$$
where $A_j$ are slowly varying in time amplitudes.

Using this represntation, we can express the driving terms in 
the equations for the
beam as follows:
\begin{gather*}
\frac{ml}{2}\ddot{\theta_j} = -\frac{ml\w^2}{2}\left(A_je^{i\w t}+A_j^*e^{-\w t}
\right)\\
\frac{ml}{2}\dot{\theta_j}^2\theta_j = -\frac{ml\w^2}{2}\left(A_j^3e^{3\w t} -
|A|_j^2A_je^{i \w t} - A_j^* |A|_j^2e^{-i\w t} + (A_j^*)^3e^{-3\w t}\right)\\
\frac{ml}{2}\ddot{\theta_j}\theta_j = -\frac{ml\w^2}{2}\left(A_j^2e^{2i\w t}
+2|A|^2_j+(A^*_j)^2e^{-2\w t}\right)\\
\frac{ml}{2}\dot{\theta}_j^2 = 
-\frac{ml\w^2}{2}\left( A_j^2e^{2i\w t} - 2|A_j|^2 +
(A_j^*)^2e^{-2i\w t}\right)
\end{gather*}
Now the response of the beam to this driving can be expressed 
via solution of the linear
equations, where the amplitudes $A$ are considered as constants:
\begin{gather*}
x(t) = \sum_j\frac{ml\w^2}{2}[ H_x(\w)A_j(1+|A|_j^2)e^{i\w t}
+H^*_x(\w)A^*_j(1+|A_j|^2)e^{-i\w t} -\\
- \left(H_x(3\w)A_j^3e^{3\w t} +  H^*_x(3\w)(A_j^*)^3e^{-3\w t} \right) ]\\
y(t) = \sum_j-ml\w^2 \left[ H_y(2\w)A^2_je^{2i\w t} + H^*_y(2\w)(A^*_j)^2
e^{-2i\w t}  \right]
\end{gather*}
and for the second derivatives we get
\begin{gather*}
\ddot{x}(t) = \sum_j-\frac{ml\w^4}{2}[ H_x(\w)A_j(1+|A|_j^2)e^{i\w t}
+H^*_x(\w)A^*_j(1+|A_j|^2)e^{-i\w t} -\\
- 9\left( H_x(3\w)A_j^3e^{3\w t} +  H^*_x(3\w)(A_j^*)^3e^{-3\w t} \right) ]\\
\ddot{y}(t) = \sum_j4ml\w^4 \left[ H_y(2\w)A^2_je^{2i\w t} + H^*_y(2\w)(A^*_j)^2
e^{-2i\w t}  \right]
\end{gather*}
Here $H_{x,y}(\w)$ are the response functions for the linear oscillators:
$$
H_{x,y}(\w) = \frac{1}{-\w^2(M+Nm)+i\gamma_{x,y}\w+k_{x,y}}
$$
Equations for the complex amplitudes $A_j(t)$ follow from the 
rewriting Eq.~(\ref{eq:aps3})
in terms of $A_j$ and averaging it over the fast time 
(basic period $2\pi/\omega$):
$$
\dot{A}_j= \frac{1}{2}A_j\left(\sigma - \rho |A_j|^2 \right) + 
\frac{1}{4i\w l}
\langle\ddot{y}\theta_je^{-i\w t}\rangle 
-\frac{1}{4i\w l}\langle\ddot{x}e^{-i\w t}\rangle
$$

After averaging only the terms with 
$\ddot{y}\theta_j\sim e^{i\w t}$ and $\ddot{x}\sim
e^{i\w t}$ survive:
$$
\dot{A_j} = \frac{1}{2}A_j\left(\sigma - \rho |A_j|^2 \right) 
+D A_j^*\sum_k A_k^2 
+S \sum_k A_k
$$
where
$$ D= -im\w^3 H_y(2\w),\qquad S= -\frac{im\w^3}{8}H_x(\w)$$ 
[here we neglected terms containing higher orders in $A_j$, due to 
smallness of the amplitudes].
Terms $\sim D$ arise from the vertical motion of the beam $\ddot{y}$,
while terms $\sim S$ are due to the horizontal motion $\ddot{x}$.

In the phase approximation we assume that the amplitudes $|A_j|$ 
are nearly constants
$|A_j|=\sqrt{\sigma/\rho}$ and the  interaction does not affect their dynamics.
Therefore for phases $\phi_j$ ($A_j = |A_j|e^{i\phi_j}$) we have the
following equations:
\begin{equation}
\dot{\phi_j} = \Omega + d\sum_k\sin(2(\phi_k-\phi_j)+\beta) +
s\sum_k\sin(\phi_k-\phi_j+\alpha)
\label{eq:app-fin}
\end{equation}
where 
 $d = \sigma\rho^{-1} |D|$, $s=|S|$, $\beta =
arg(D)$ and $\alpha = arg(S)$.
The freqeuncy is determined as 
$\Omega =Im\left( \sigma\rho^{-1} D+S \right)$. The obtained system 
is the Kuramoto model with bi-harmonic coupling.

\end{document}